\documentclass[a4paper,fleqn,usenatbib]{mnras}
\usepackage{graphicx}
\usepackage[T1]{fontenc}
\usepackage{ae,aecompl}

\usepackage{amsmath}	
\usepackage{amssymb}	

\usepackage{color}

\hypersetup{draft}

\title[The compact triple stars TICs 167692429 and 220397947]{TICs 167692429 and 220397947: The first compact hierarchical triple stars discovered with {\em TESS}}

\author[Borkovits et al.]{
T.~Borkovits$^{1,2}$\thanks{E-mail: borko@electra.bajaobs.hu}, 
S.~A.~Rappaport$^3$,
T.~Hajdu$^{2,4}$, 
P.~F.~L.~Maxted$^5$,
A. P\'al$^{2}$,
\newauthor
E.~Forg\'acs-Dajka$^6$,
P.~Klagyivik$^{7}$,
T.~Mitnyan$^{1}$\\ 
$^1$ Baja Astronomical Observatory of Szeged University, H-6500 Baja, Szegedi \'ut, Kt. 766, Hungary \\
$^2$ Konkoly Observatory, Research Centre for Astronomy and Earth Sciences, \\
 H-1121 Budapest, Konkoly Thege Miklós \'ut 15-17, Hungary \\
$^3$ Department of Physics, Kavli Institute for Astrophysics and Space Research, M.I.T., Cambridge, MA 02139, USA\\
$^4$ MTA CSFK Lend{\"u}let Near-Field Cosmology Research Group \\
$^5$ Astrophysics Group, Keele University, Staffordshire, ST5 5BG, UK\\
$^6$ E\"otv\"os Lor\'and University, Department of Astronomy, H-1118 Budapest, P\'azm\'any P\'eter stny. 1/A, Hungary\\
$^7$ Deutsches Zentrum f\"ur Luft und Raumfahrt, Institut f\"ur Planetenforschung Rutherfordstrasse 2, 12489 Berlin, Germany, DE}

\begin{document}

\date{-}

\pagerange{\pageref{firstpage}--\pageref{lastpage}} \pubyear{2019}

\maketitle

\label{firstpage}

\begin{abstract}
We report the discovery and complex analyses of the first two compact hierarchical triple star systems discovered with {\em TESS} in or near its southern continuous viewing zone during Year 1.  Both TICs\,167692429 and 220397947 were previously unknown eclipsing binaries, and the presence of a third companion star was inferred from eclipse timing variations exhibiting signatures of strong 3rd-body perturbations and, in the first system, also from eclipse depth variations.  We carried out comprehensive analyses, including the simultaneous photodynamical modelling of {\em TESS} and archival ground-based WASP lightcurves, as well as eclipse timing variation curves.  Also, for the first time, we included in the simultaneous fits multiple star spectral energy distribution data and theoretical PARSEC stellar isochrones, taking into account Gaia DR2 parallaxes and cataloged metallicities.   We find that both systems have twin F-star binaries and a lower mass tertiary star.  In the TIC\,167692429 system the inner binary is moderately inclined ($i_\mathrm{mut}=27\degr$) with respect to the outer orbit, and the binary vs. outer (triple) orbital periods are 10.3 vs.~331 days, respectively.  The mutually inclined orbits cause a driven precession of the binary orbital plane which leads to the disappearance of binary eclipses for long intervals.  In the case of TIC\,220397947 the two orbital planes are more nearly aligned and the inner vs.~outer orbital periods are 3.5 and 77 days, respectively.  In the absence of radial velocity observations, we were unable to calculate highly accurate masses and ages for the two systems. According to stellar isochrones TIC\,167692429 might be either a pre-main sequence or an older post-MS system. In the case of TIC\,220397947 our solution prefers a young, pre-MS scenario. 
\end{abstract}

\begin{keywords}
binaries: eclipsing -- stars: individual: 
\end{keywords}


\section{Introduction}
\label{sec:intro}

Close, compact, hierarchical, multiple stellar systems, i.e., multiples at the lowest end of the outer period domain, comprise a small but continuously growing group of the triple and multiple star zoo. The significance of these most compact systems lies in the fact that they challenge or, at least probe, the alternative multiple star formation scenarios by their extreme properties. Furthermore, due to the relatively readily observable short-term dynamical interactions amongst the components, their dynamical as well as astrophysical properties can be explored with a high precision. For example, one key parameter that can be measured through the dynamical interactions in a compact triple system is the mutual inclination of the inner and outer orbits. This quantity is expected to be a primary tracer of the formation process(es) of triples and their later dynamical evolutions leading to the present-day configurations of the systems \citep[see, e.\,g.][and further references therein]{fabryckytremaine07,moekratter18,tokovininmoe19}.  Other, less emphasized parameters which can be deduced almost exclusively from the observations of short-term dynamical interactions of such systems are the orientations of the orbits relative to the intersections of their orbital planes \citep{borkovitsetal11}. These parameters, the so-called dynamical arguments of periastron (i.e., the argument of periastron measured from the ascending node of the respective orbit relative to the invariable plane of the system instead of the tangential plane of the sky) have substantial importance for the long-term dynamical evolution of highly inclined multiples \citep[see, e.\,g.][and further references thereins]{fordetal00,naoz16}.

Before the advent of the era of space telescopes dedicated to searches for transiting extrasolar planets, only a very limited number of extremely compact triple or multiple stars were known. The preferred method for finding close tertiary components, before space missions, was the radial velocity (RV) measurements of known close binaries (discovered either by photometry or spectroscopy). Third stellar components orbiting eclipsing binaries (EB) could also be detected photometrically (and, frequently were), through eclipse timing variations (ETVs) due either to the light-travel time effect (LTTE) or to direct third-body perturbations.  Ground-based detection of {\em close} third stellar companions via ETVs, however, is less efficient for the following reasons. First, the LTTE is biased toward longer periods and more massive tertiaries since the amplitude of an LTTE-caused ETV is $A_\mathrm{LTTE}\propto\frac{m_\mathrm{C}}{m_\mathrm{ABC}^{2/3}}P_2^{2/3}$. For the shortest outer period systems this usually remains below the detection limit of ETVs found in strongly inhomogeneous ground-based eclipse timing observations. Second, the amplitude of the short-term dynamical perturbations on the ETVs scales with both the inner period and the inner to outer period ratio ($A_\mathrm{dyn}\propto P_1^2/P_2$), therefore, it becomes observable at the accuracy of ground-based measurements only for longer period EBs, which are  unfavoured for ground-based photometry \citep[see, the discussion of][]{borkovitsetal03}.

While today's dedicated spectroscopic surveys lead continually to the discovery of new very compact multiple stellar systems \citep[see, e.\,g., most recently][and further references therein]{tokovinin19}, in the era of space photometry, photometric detection of these systems has became the dominant discovery mode.  This breakthrough was largely due to the {\em Kepler} space telescope \citep{boruckietal10} thanks to which the number of the closest compact triple (and in part, probably multiple) stellar systems has grown significantly over the last decade. \citet{Borkovits2016} have identified more than 200 triple star candidates amongst the $\sim 2900$ EBs \citep{kirketal16} observed quasi-continuously by {\em Kepler} during its four-year-long primary mission. Eight of these triple candidates have outer periods less than 100 days and an additional $\sim 27$ systems were detected with outer periods less than 1 year. Moreover, \citet{Borkovits2016} have shown that the absence of further very short outer period triples amongst {\em Kepler}'s EBs cannot be an observational selection effect.  At least three additional very short outer period triple stars were detected in the fields of the {\em K2} mission; two of them, HD\,144548 \citep{alonsoetal15} and EPIC\,249432662 \citep{Borkovits2019a} exhibit outer eclipses, while the third, HIP\,41431, which was discovered independently as a spectroscopic triple, is indeed at least a 2+1+1 quadruple system \citep{Borkovits2019b}. Moreover, \citet{hajduetal17} identified four triple-star candidates with outer periods probably less than 1 year amongst EBs observed by the CoRoT spacecraft \citep{auvergneetal09}.

In this paper we report the discovery and detailed analysis of the first two close, compact, hierarchical triple (or multiple) stellar systems, TICs\,167692429 and 220397947 observed by the {\em TESS} spacecraft \citep{ricker15}. Both systems consist of previously unknown EBs composed of nearly equal mass ($q_1>0.9$) F-type stars (in the parlance of the binary- and multiple-star community: `solar-type stars').  Both EBs have a detached configuration, and the orbital periods are $P_1=10.26$ and $P_1=3.55$\,days for TICs\,167692429 and 220397947, respectively. Both of them exhibit three-body perturbation-dominated, short-term ETVs, with periods $P_2=331$ days for TIC\,167692429 and $P_2=77$ days for TIC\,220397947.  For TIC\,167692429 the moderately eccentric EB exhibits eclipse depth variations with a clear signature of an outer periastron passage bump as a sign of an inclined eccentric tertiary. Both systems were observed with the WASP-South cameras \citep{hellieretal11} between 2008 and 2014. These early lightcurves have been included into our analyses of the two systems.  On the other hand, however, no RV measurements are available for these systems. Therefore, we use spectral energy distribution (`SED') information, Gaia DR2 parallaxes, and theoretical PARSEC isochrones to constrain stellar masses and temperatures throughout the joint photodynamical analysis of the {\em TESS} and WASP lightcurves and ETV curves.

In Section 2 we describes all the available observational data, as well as their preparation for the analysis. Then, Section 3 provides a full explanation of the steps of the joint physical and dynamical modeling of the light- and ETV curves, SED, parallax and stellar isochrones. In Sections 4 and 5 we discuss the results from astrophysical and dynamical points of views. Finally, in Sect. 6 we draw conclusions from our findings.



\section{Observational data}
\label{sec:obs}

\begin{table*}
\centering
\caption{Main characteristics of TICs 167692429 and 220397947}
\label{tab:object}
\medskip
\begin{tabular}{lr l  l}
\hline
Parameter & &\multicolumn{2}{c}{Value} \\
\hline
Identifiers & TIC & 167692429 & 220397947 \\
            & TYC & 8899-18-1 & 8515-663-1\\
            &2MASS& 06505184-6325519 & 04360354-5804334\\
Position &(J2015.5, Gaia DR2) & 06:50:51.852, -63:25:51.76 & 04:36:03.537, -58:04:33.09\\
PM & $\mu_\alpha$, $\mu_\delta$ (mas~yr$^{-1}$, Tycho-2)  & +4.6$\pm$3.4, +16.1$\pm$3.2 & -4.1$\pm$2.5, +28.2$\pm$2.3 \\
PM & $\mu_\alpha$, $\mu_\delta$ (mas~yr$^{-1}$, Gaia DR2) & +1.82$\pm$0.06, +15.99$\pm$0.05 & +0.97$\pm$0.04, +24.46$\pm$0.05 \\
Parallax & (mas, Gaia DR2) & 1.41 $\pm$ 0.03 & 2.87 $\pm$ 0.02\\
$T_\mathrm{eff}$ & (K, TIC-8) & $6474\pm117$ & $6257\pm131$ \\
                 & (K, Gaia DR2) & $6342_{-228}^{+220}$ & $6289_{-158}^{+238}$\\
$\log g$ & (cgs, TIC-8) & $3.48\pm0.09$ & $4.02\pm0.09$\\
Metallicity & $[M/H]$ & $-0.310\pm0.046$ & $-0.669\pm0.061$ \\
Optical photometry$^a$& $B$, $V$ (mag) & $11.408(21)$, $10.931(47)$ & 11.301(28), 10.846(67)\\
                   & $g'$, $r'$, $i'$ (mag) & $...$, $10.875(42)$, $10.784(33)$ & 11.050(88), 10.793(41), 10.757(62)\\
TYCHO-2 photometry$^b$& $B_T$, $V_T$ (mag) & 11.491(64), 10.928(61) & 11.585(76), 10.949(68)\\
Gaia photometry    & $G$, $B_P$, $R_P$ (mag) & $10.8482(3)$, $11.1074(7)$, $10.4571(5)$ & 10.7535(5), 11.0198(13), 10.3572(9)\\
Infrared photometry$^c$& $J$, $H$, $K_s$ (mag) & $10.014(22)$, $9.805(21)$, $9.764(23)$ & 9.913(23), 9.700(26), 9.598(20) \\
WISE photometry$^d$ & $w_1$, $w_2$ (mag) & $9.723(23)$, $9.753(20)$ &  9.567(21), 9.582(17)\\
Extinction & $E(B-V)$ (mag) & $0.064\pm0.010$ & $0.008\pm0.010$\\
\hline
\end{tabular}

{\em Notes.} Sources of the SED information. $a$: AAVSO Photometric All Sky Survey (APASS) DR9, \citep{APASS}, \url{http://vizier.u-strasbg.fr/viz-bin/VizieR?-source=II/336/apass9}; $b$:Tycho-2 catalogue \citep{TYCHO2}; $c:$ 2MASS catalogue \citep{2MASS}; $d$: WISE point source catalogue \citep{WISE}.  All the other data are taken directly from the Gaia DR2 \citep{GaiaDR2} and TIC-8 \citep{TIC} catalogs. The original sources are listed in Sect.\,\ref{Sect:SEDGaia}. 
\end{table*}

\subsection{{\em TESS} observations}

\subsubsection{TIC\,167692429}

\begin{figure*}
\begin{center}
\includegraphics[width=0.49 \textwidth]{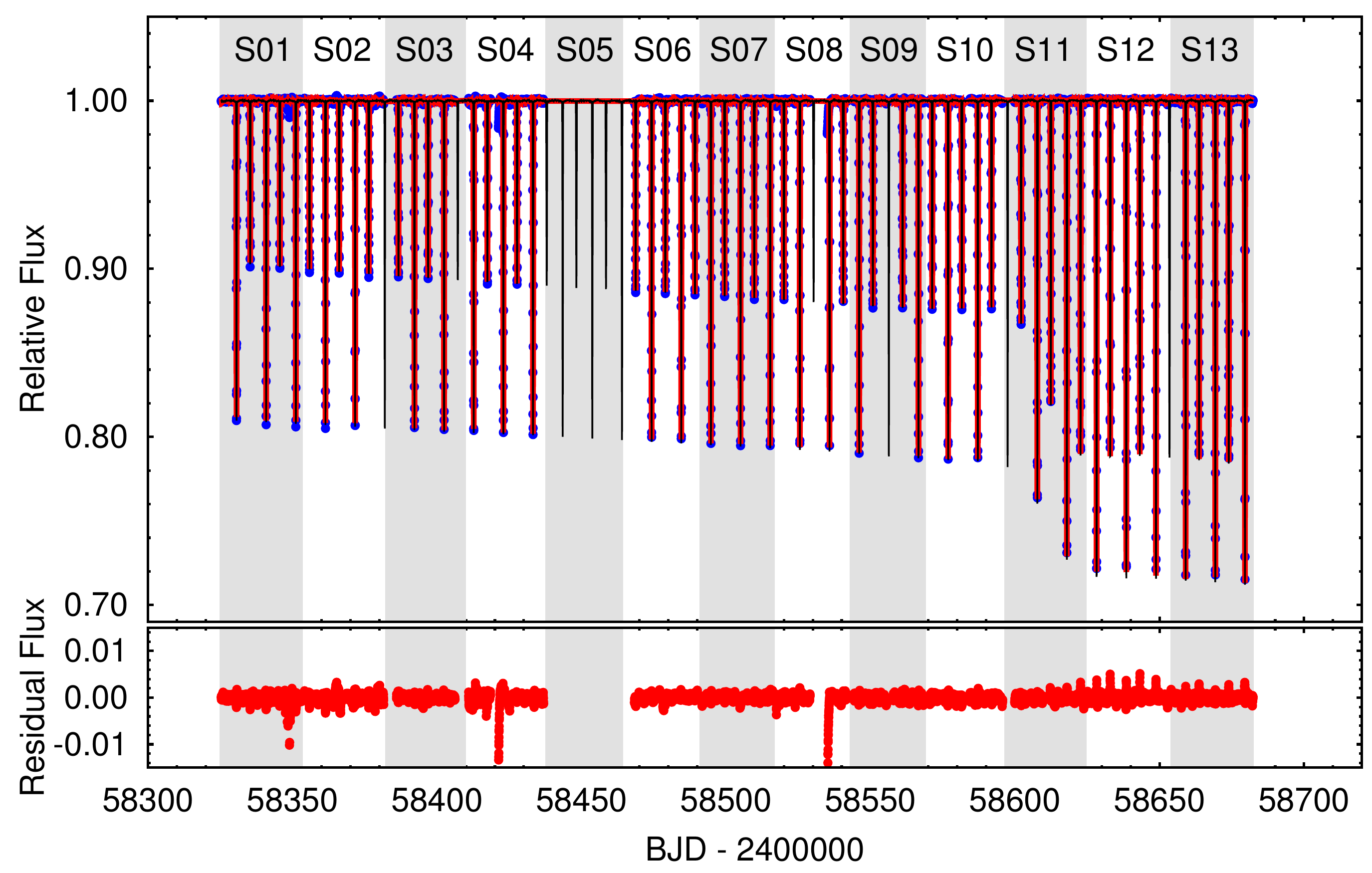}
\includegraphics[width=0.49 \textwidth]{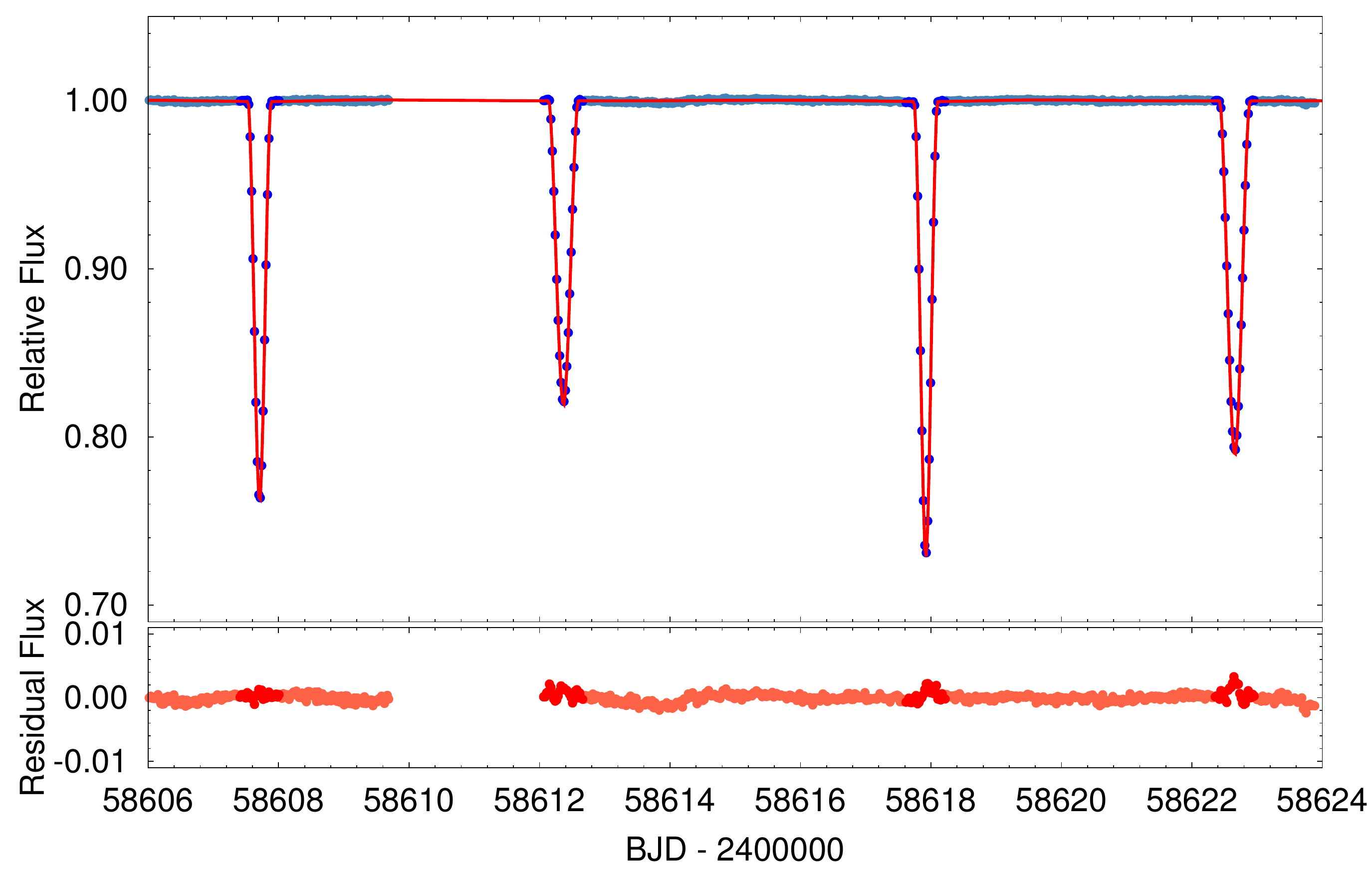}
\caption{ The {\em TESS} lightcurve of TIC\,167692429. Instead of the full resolution detrended SAP SC flux curve, we plot the 1800-sec binned lightcurve which was used for the photodynamical analysis (see text for details). {\em First row, upper panel}: The complete (Sectors 1--4, 6--13) lightcurve is plotted with blue dots. Red curve represents the cadence-time corrected photodynamical model solution, see Sect.\,\ref{sec:dyn}. The thin black curves in the data gaps show the continuously sampled model lightcurve. Alternating gray and white stripes denote the consecutive {\em TESS} sectors. {\em Second row, upper panel:} An 18-day-long section of the lightcurve around the time of periastron passage of the third star. The dark blue circles in the $\pm0\fp3$ phase-domain around each individual minimum represents the 1800-sec binned flux values used for the photodynamical model, while the other out-of-eclipse data (not used in the modelling) are plotted as grey circles. The red curve is the cadence-time corrected photodynamical model solution; the residuals to the model are also shown in the bottom panels.}
\label{fig:eclipsefitT167} 
\end{center}
\end{figure*}

\begin{figure*}
\begin{center}
\includegraphics[width=0.49 \textwidth]{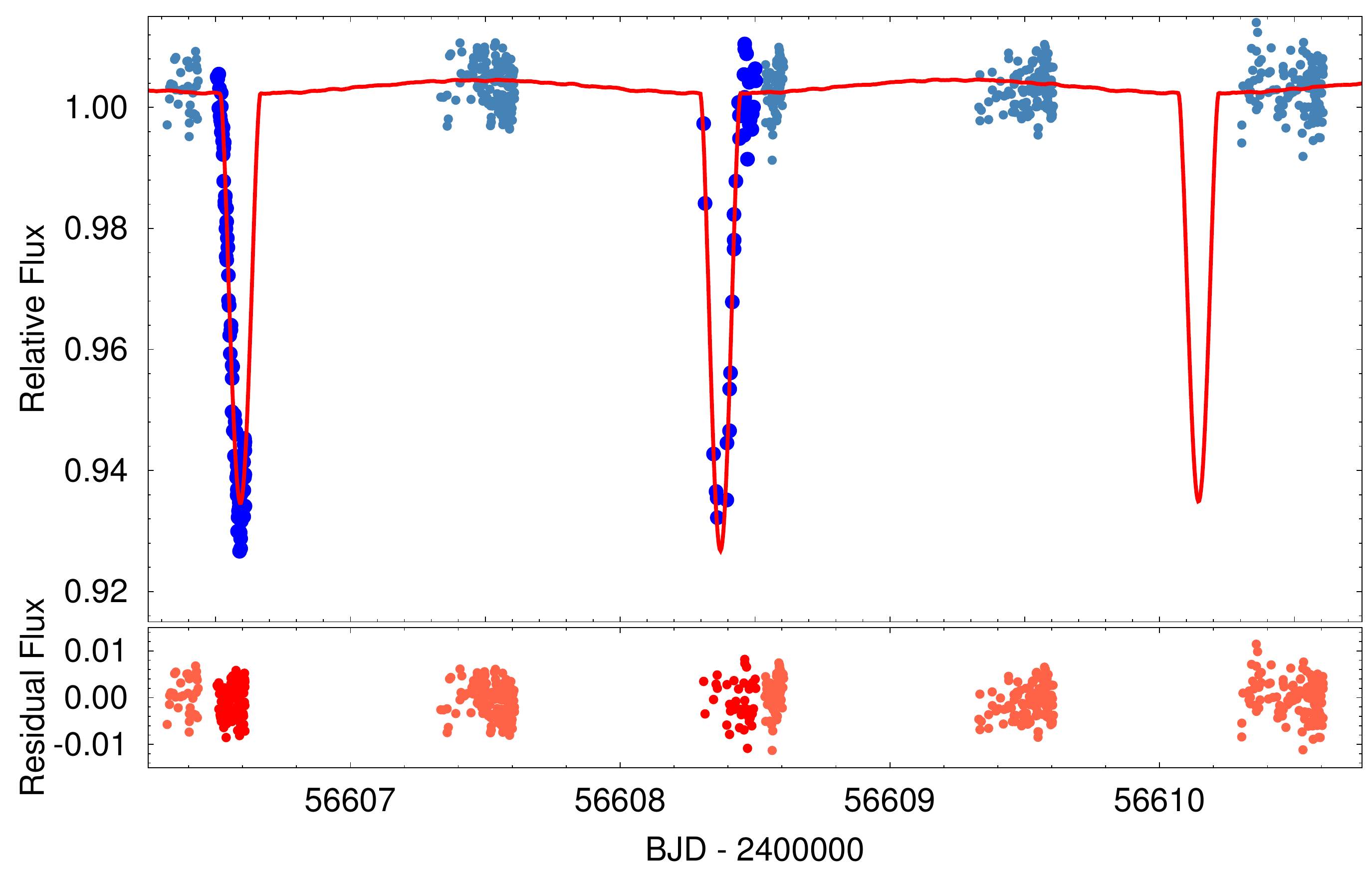}
\includegraphics[width=0.49 \textwidth]{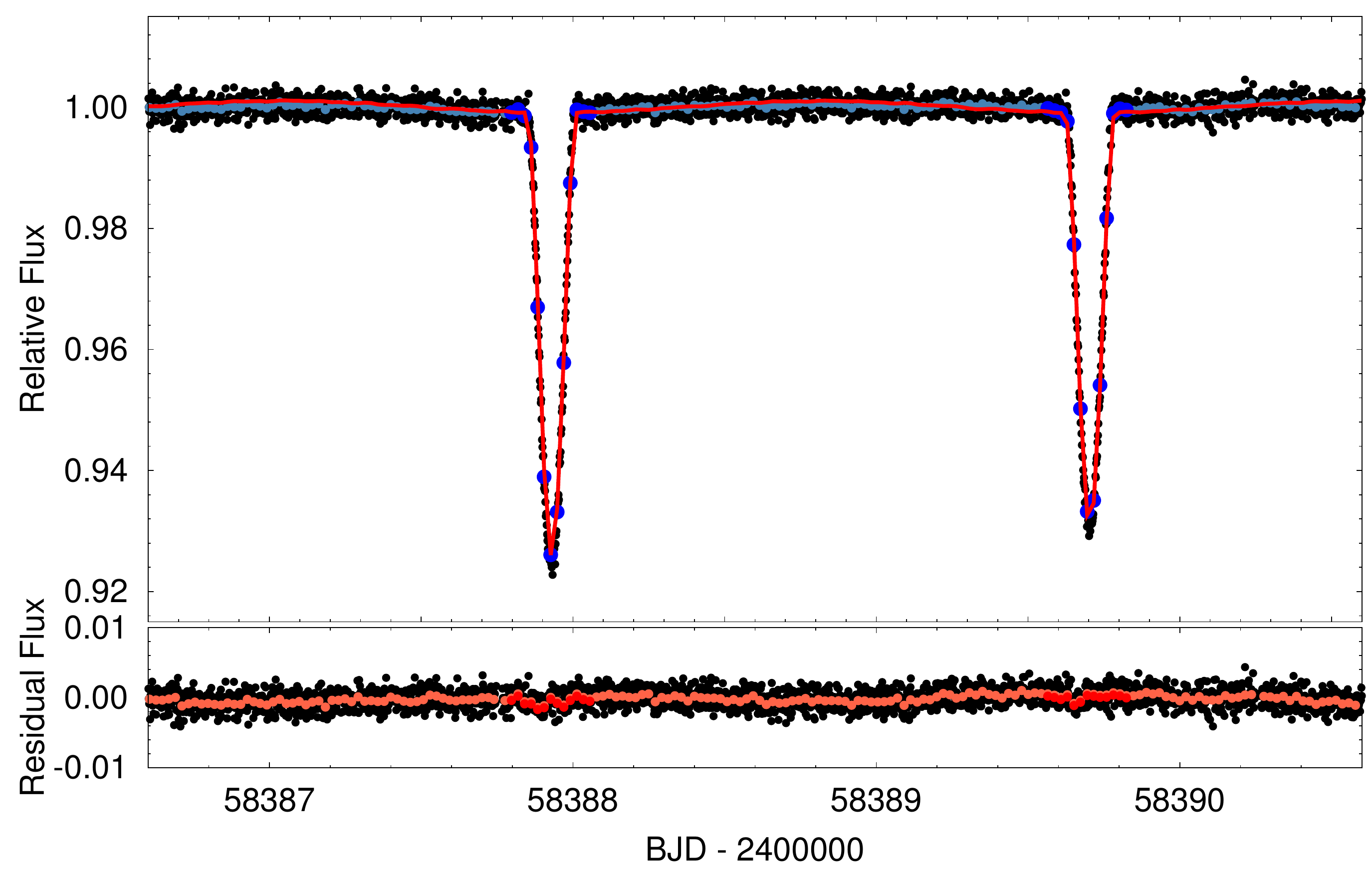}
\caption{ {\em TESS} and WASP lightcurves of TIC\,220397947. {\em Upper left panel}: A 4.5-day-long section of SWASP lightcurves with the photodynamical model solution. Dark blue circles show those observations which were used for photodynamical modelling, while the other, unmodelled, observations are plotted with light blue. Furthermore, the red curve represents the photodynamical model solution. {\em Upper right panel:} A 4-day-long section of the lightcurve obtained by {\em TESS} during Sector 1 observations. Black dots represent the PDC-SAP short cadence fluxes. The dark blue circles in the $\pm0\fp04$ phase-domain around each individual minimum represent the 1800-sec binned flux values used for the photodynamical model, while the other, similarly binned, out-of-eclipse data (not used for the modelling) are plotted by light blue circles. The red curve is the cadence-time corrected photodynamical model solution; the residuals to the model are also shown in the bottom panels.}
\label{fig:eclipsefitT220} 
\end{center}
\end{figure*}

\begin{figure*}
\centerline{
\includegraphics[width=16cm]{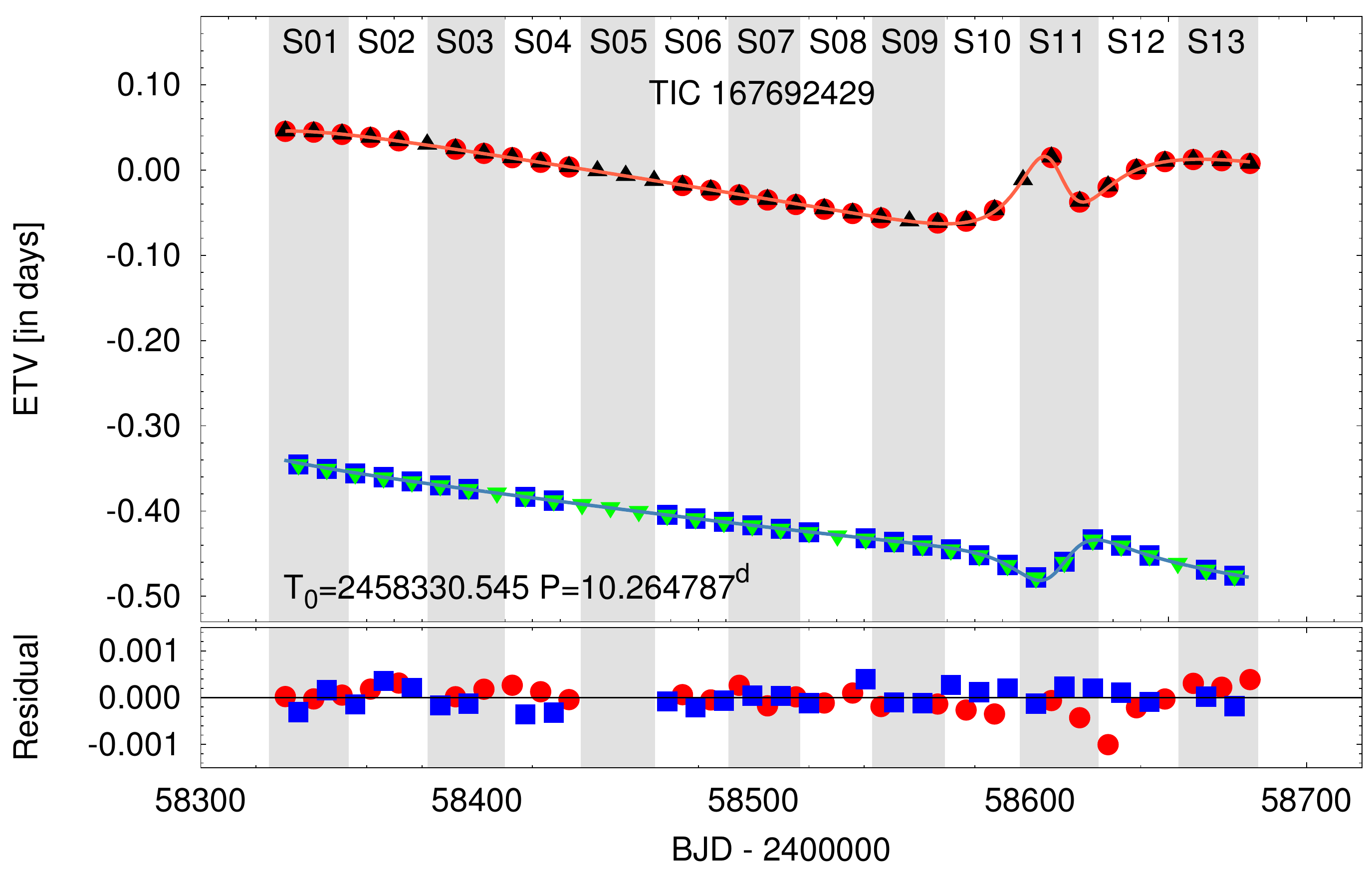}
}
\caption{ Eclipse timing variations of TIC\,167692429. Red circles and blue boxes represent the primary and secondary ETVs, respectively, calculated from the observed eclipse events, while black upward and green downward triangles show the corresponding primary and secondary ETV, determined from the photodynamical model solution. Furthermore, orange and light blue lines represent approximate analytical ETV models for the primary and secondary eclipses. The residuals of the observed vs photodynamically modelled ETVs are plotted in the bottom panel. As before, gray and white stripes denote the consecutive {\em TESS} sectors.}
\label{fig:ETV}
\end{figure*}

TIC\,167692429 was observed by the {\em TESS} spacecraft \citep{ricker15} during Year 1 in short cadence (SC) mode nearly continuously with the exception of Sector 5. Similar to other EBs in (or near) the continuous viewing zone (`CVZ') we downloaded the calibrated SC data files for each sector from the MAST Portal\footnote{\url{https://mast.stsci.edu/portal/Mashup/Clients/Mast/Portal.html}}.  The {\em TESS} lightcurve of TIC\,167692429 is presented in Fig.\,\ref{fig:eclipsefitT167}.  

As soon as the data from the first four sectors became available and were downloaded, we realized that eclipse timing variations (`ETVs') of the primary and secondary eclipses exhibit non-linear, and mostly anticorrelated behaviour that are most probably of dynamical origin (see, e.g., \citealt{Borkovits2015}). Data from this object over the next few sectors indicated that the eclipse depths were slightly increasing, and that the primary and secondary ETVs, though converging weakly, did not show any characteristic non-linearity up to the second half of Sector 10. Then in Sector 11 both the eclipse depths and their timings changed dramatically hinting at the periastron passage of a third body in a significantly inclined and remarkably eccentric orbit.  

This led us to collect all the observations and then carry out a complex photodynamical lightcurve and ETV analysis (see, e.g., \citealt{Borkovits2019a}). For this purpose we used the Simple Aperture Photometry (`SAP') data. We removed all data points flagged with non-zero values. Then we normalized the fluxes from each sector and concatenated them. Finally, we used the software package {\sc W{\={o}}tan} \citep{hippkeetal19} to detrend the lightcurves removing the instrumental effects. In order to check that we did not remove those lightcurve features which might have arisen from binary star interactions (e.g., ellipsoidal variations, reflection effect, Doppler-boosting) we made phase-folded, binned and averaged lightcurves both from the original and the detrended lightcurves, and adjusted the {\sc W{\={o}}tan} flattening parameter to such a value that the folded, binned, averaged detrended lightcurves qualitatively preserve the same out-of-eclipse features as the non-detrended ones.

\subsubsection{TIC\,220397947}

This target was observed by {\em TESS} during Sectors 2--6, 8, 9 and 12. Short cadence lightcurves are available only from Sectors 3, 4, 9 and 12. Similar to the other target above, SC lightcurves were downloaded from MAST Portal. For the long cadence (LC) lightcurves of those sectors where SC data were not available we processed the {\em TESS} full-frame images using a convolution-based differential photometric pipeline, based on the various tasks of the {\sc Fitsh} package \citep{2012MNRAS.421.1825P}. Namely, small stamps with a size of $64\times64$ pixels were extracted centered on the target source and a combined, stray light-free median image (created from 11 individual frames) was used as a reference for the image subtraction algorithm.  The implemented image subtraction algorithm also accounts for the variations in the point-spread functions (PSF) by fitting the appropriate convolution transformation. While the actual variations in PSF are comparatively small, this step is important for removing the effect of the gradual drift in the light centroid positions caused by the differential velocity aberration.  Instrumental fluxes were obtained using the appropriate equations provided by \cite{2009PhDT.........2P}, while the zero-point reference was computed using the Gaia DR2 $R_P$ magnitudes \citep{2016A&A...595A...1G,GaiaDR2}. This $R_P$ magnitude is a rather accurate estimation due to the significant overlap of the {\em TESS} and Gaia passbands \citep{ricker15,2010A&A...523A..48J}. We also downloaded LC lightcurves for Sectors 2--5 from the TESS Full Frame Image Portal\footnote{\url{https://filtergraph.com/tess_ffi/}} which hosts the data products from the pipeline of \citet{oelkersstassun18}. While for the determination of the mid-eclipse times for each eclipse observed by {\em TESS} we used both the long and the short cadence data, for the photodynamical analysis we used only the {\sc W{\={o}}tan}-detrended short cadence SAP lightcurve.  A segment of the {\em TESS} lightcurve of TIC\,220397947 is presented in the right panel of Fig.\,\ref{fig:eclipsefitT220}.

\subsection{WASP observations}

Both TICs\,167692429 and 220397947 are amongst the millions of stars that have been observed as part of the WASP survey. The survey is described in \citet{2006PASP..118.1407P} and \citet{2006MNRAS.373..799C}. The WASP instruments each consist of an array of 8 cameras with Canon 200-mm f/1.8 lenses and  2k$\times$2k $e$2$V$ CCD detectors providing images with a field-of-view  of $7.8^{\circ}\times 7.8^{\circ}$ at an image scale of 13.7 arcsec/pixel. Images are obtained through a broad-band filter covering 400-700\,nm.  From July 2012 the WASP-South instrument was operated using 85-mm, f/1.2 lenses and an r$^{\prime}$ filter. With these lenses the image scale is 33 arcsec/pixel. Fluxes are measured in an aperture with a radius of 48 arcsec for the 200-mm data, and 132 arcsec for the 85-mm data.  The data are processed with the SYSRem algorithm \citep{2005MNRAS.356.1466T} to remove instrumental effects.

Observations of TIC\,167692429 were obtained simultaneously in two cameras on WASP-South over four observing seasons, from 2008 Sep 29 to 2012 Mar 23. 
During the four seasons of WASP observations the target did not exhibit any eclipses, however, there is a clear dip in the measured flux with a depth of about 5.4\% and a duration of at least 1.17 days. There are no other dips of comparable depth and width in the WASP data.

Observations of TIC~220397947 were obtained simultaneously in two cameras on WASP-South over four observing seasons, from 2010 Aug 05 to 2014 Dec 19. A segment of the WASP lightcurve for TIC\,220397947 is shown in the left panel of Fig.\,\ref{fig:eclipsefitT220}.

\subsection{ETV data}

We determined the mid-time of each eclipse observed by {\em TESS} using both the SC and LC lightcurves, though for all further analyses, eclipse times obtained from the LC lightcurves were used only where SC data were unavailable. The method we used is described in detail by \citet{Borkovits2016}. Note, that in the case of TIC\,167692429, for the eclipse depth and duration variations during the 11 months of the {\em TESS} observations, in addition to the template fitting approach \citep{Borkovits2016}, we also found the eclipse times by fitting a parabola to each eclipse bottom. The two methods, however, resulted in very similar values, well within the estimated accuracies; therefore, we decided to use the first set of the ETV data obtained by using the method of \citet{Borkovits2016}.

Regarding the WASP observations, TIC\,167692429 did not exhibit eclipses during these measurements. By contrast, for TIC\,220397947 several eclipses were observed during the four seasons of the WASP observations. Most of these eclipses, however, were unfortunately only partially covered and therefore, they do not lead to accurate eclipse timing determinations. Instead of the determination of the minima from a few average seasonal lightcurves, we found it to be more appropriate for our purposes to select the relatively better observed individual eclipses and determine their mid-eclipse times. Though these times of minima exhibit large scatter, they manifest a significant trend which leads us to the conclusion that TIC\,220397947 might indeed be a 2+1+1 type quadruple system.    

\begin{figure*}
\centerline{
\includegraphics[width=0.49\textwidth]{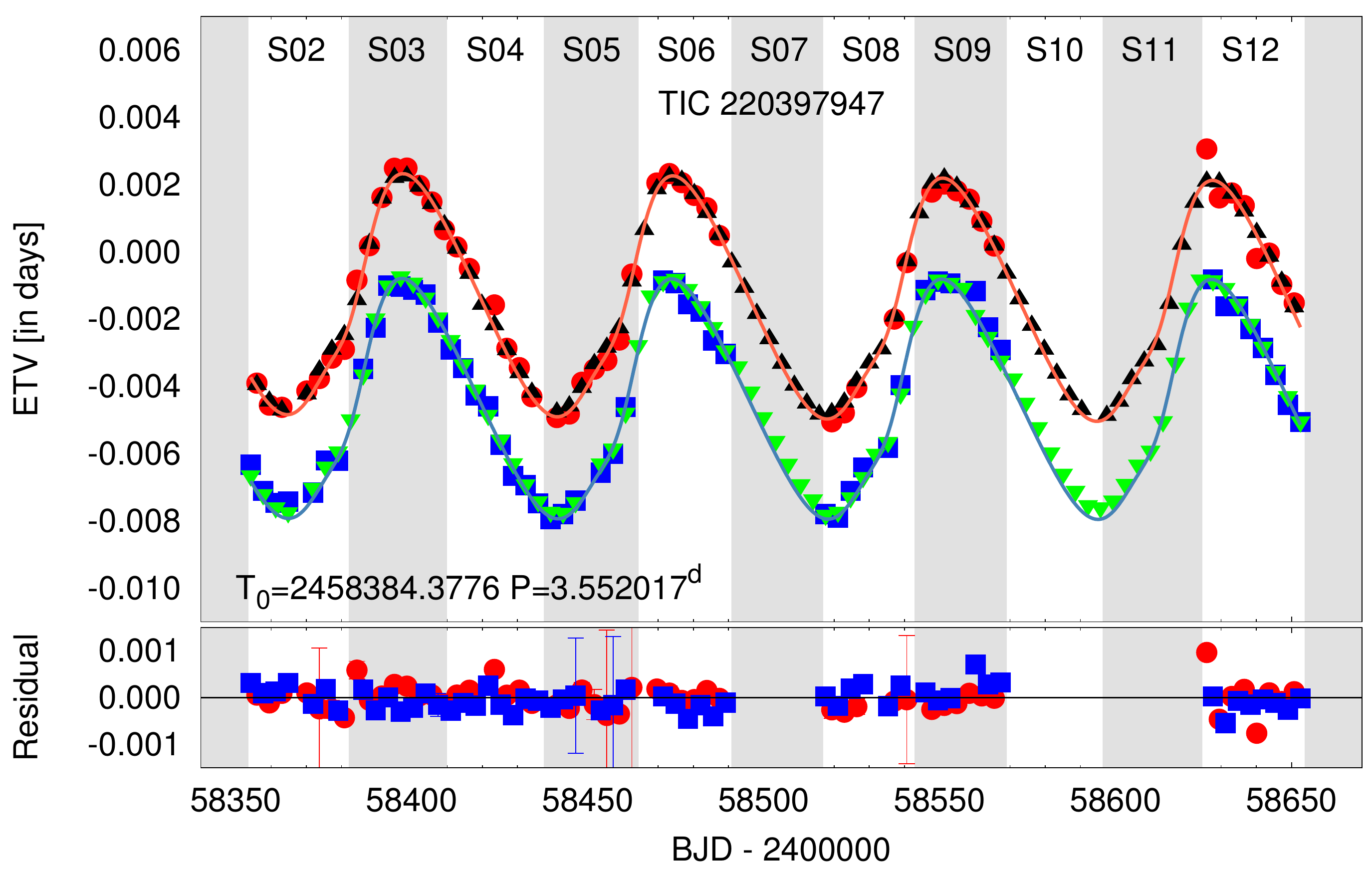}\includegraphics[width=0.49\textwidth]{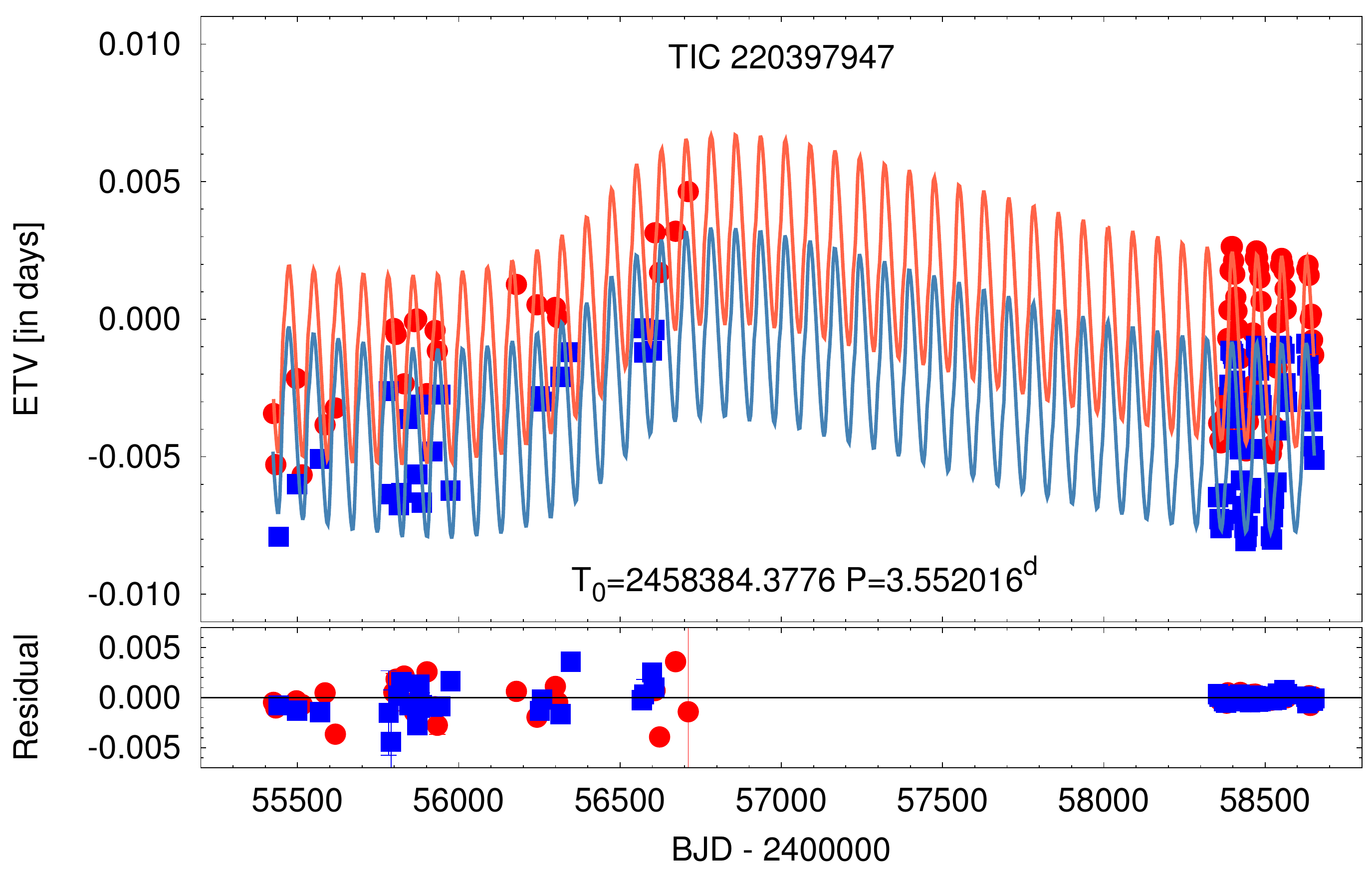}
}
\caption{{\it Left panel:} Eclipse timing variations of TIC\,220397947 during the 11 months of {\em TESS} observations. Red circles and blue boxes represent the primary and secondary ETVs, respectively, calculated from the observed eclipse events, while black upward and green downward triangles show the corresponding primary and secondary ETV, determined from the photodynamical model solution. Furthermore, orange and light blue lines represent approximate analytical ETV models for the primary and secondary eclipses. {\it Right panel:} ETVs of TIC\,220397947 from the beginning of the SWASP observations. As before, red circles and blue boxes represent the primary and secondary ETVs, but here orange and light blue lines connect the ETV points determined from the four-body dynamical modelling. The residuals of the observed vs photodynamically modelled ETVs are plotted in the bottom panels.}
\label{fig:ETV2}
\end{figure*}

The times of minima of the two systems are listed in Tables\,\ref{Tab:TIC167692429_ToM} and \ref{Tab:TIC220397947_ToM}, while the ETV curves are shown in Figs.\,\ref{fig:ETV} and \ref{fig:ETV2} for TICs\,167692429 and 22039794, respectively. 

\begin{table*}
\caption{Times of minima of TIC\,167692429. }
 \label{Tab:TIC167692429_ToM}
\begin{tabular}{@{}lrllrllrl}
\hline
BJD & Cycle  & std. dev. & BJD & Cycle  & std. dev. & BJD & Cycle  & std. dev. \\ 
$-2\,400\,000$ & no. &   \multicolumn{1}{c}{$(d)$} & $-2\,400\,000$ & no. &   \multicolumn{1}{c}{$(d)$} & $-2\,400\,000$ & no. &   \multicolumn{1}{c}{$(d)$} \\ 
\hline
58330.590372 &    0.0 & 0.000094 & 58468.715114 &   13.5 & 0.000075 & 58576.839937 &   24.0 & 0.000095 \\ 
58335.331948 &    0.5 & 0.000079 & 58474.233865 &   14.0 & 0.000094 & 58581.580405 &   24.5 & 0.000074 \\ 
58340.854210 &    1.0 & 0.000094 & 58478.975608 &   14.5 & 0.000080 & 58587.117373 &   25.0 & 0.000095 \\ 
58345.591623 &    1.5 & 0.000083 & 58484.492996 &   15.0 & 0.000093 & 58591.833896 &   25.5 & 0.000067 \\ 
58351.116444 &    2.0 & 0.000098 & 58489.236426 &   15.5 & 0.000065 & 58602.083879 &   26.5 & 0.000066 \\ 
58355.851013 &    2.5 & 0.000088 & 58494.752538 &   16.0 & 0.000081 & 58607.708972 &   27.0 & 0.000087 \\ 
58361.377729 &    3.0 & 0.000085 & 58499.497258 &   16.5 & 0.000066 & 58612.367185 &   27.5 & 0.000083 \\ 
58366.111482 &    3.5 & 0.000097 & 58505.011331 &   17.0 & 0.000085 & 58617.921526 &   28.0 & 0.000097 \\ 
58371.638409 &    4.0 & 0.000086 & 58509.758041 &   17.5 & 0.000067 & 58622.658055 &   28.5 & 0.000063 \\ 
58376.371468 &    4.5 & 0.000073 & 58515.270790 &   18.0 & 0.000098 & 58628.203675 &   29.0 & 0.000179 \\ 
58386.631345 &    5.5 & 0.000069 & 58520.018732 &   18.5 & 0.000070 & 58632.915717 &   29.5 & 0.000067 \\ 
58392.158197 &    6.0 & 0.000095 & 58525.530013 &   19.0 & 0.000091 & 58638.489436 &   30.0 & 0.000103 \\ 
58396.891705 &    6.5 & 0.000073 & 58535.789778 &   20.0 & 0.000106 & 58643.168928 &   30.5 & 0.000071 \\ 
58402.418095 &    7.0 & 0.000084 & 58540.541099 &   20.5 & 0.000060 & 58648.763268 &   31.0 & 0.000108 \\ 
58412.677795 &    8.0 & 0.000098 & 58546.049450 &   21.0 & 0.000099 & 58659.030530 &   32.0 & 0.000105 \\ 
58417.412264 &    8.5 & 0.000072 & 58550.801516 &   21.5 & 0.000068 & 58663.681664 &   32.5 & 0.000070 \\ 
58422.937174 &    9.0 & 0.000099 & 58561.062193 &   22.5 & 0.000075 & 58669.293862 &   33.0 & 0.000114 \\ 
58427.672749 &    9.5 & 0.000082 & 58566.573032 &   23.0 & 0.000096 & 58673.939620 &   33.5 & 0.000058 \\ 
58433.196453 &   10.0 & 0.000080 & 58571.322567 &   23.5 & 0.000069 & 58679.555548 &   34.0 & 0.000117 \\ 
\hline
\end{tabular}

{\em Notes.} Integer and half-integer cycle numbers refer to primary and secondary eclipses, respectively.
\end{table*}
%

\begin{table*}
\caption{Times of minima of TIC\,220397947}
 \label{Tab:TIC220397947_ToM}
\begin{tabular}{@{}lrllrllrl}
\hline
BJD & Cycle  & std. dev. & BJD & Cycle  & std. dev. & BJD & Cycle  & std. dev. \\ 
$-2\,400\,000$ & no. &   \multicolumn{1}{c}{$(d)$} & $-2\,400\,000$ & no. &   \multicolumn{1}{c}{$(d)$} & $-2\,400\,000$ & no. &   \multicolumn{1}{c}{$(d)$} \\ 
\hline
55425.544843 & -833.0 & 0.000173 & 58371.938241 &   -3.5 & 0.000135 & 58476.732255 &   26.0 & 0.000062 \\ 
55432.647018 & -831.0 & 0.000425 & 58373.717958 &   -3.0 & 0.000110 & 58478.504346 &   26.5 & 0.000148 \\ 
55441.524433 & -828.5 & 0.000225 & 58375.491232 &   -2.5 & 0.000046 & 58480.283904 &   27.0 & 0.000126 \\ 
55496.586439 & -813.0 & 0.000940 & 58377.270547 &   -2.0 & 0.000234 & 58482.056165 &   27.5 & 0.000180 \\ 
55498.358609 & -812.5 & 0.000263 & 58379.043220 &   -1.5 & 0.000111 & 58483.835542 &   28.0 & 0.001335 \\ 
55514.343015 & -808.0 & 0.000196 & 58380.822856 &   -1.0 & 0.000062 & 58485.607303 &   28.5 & 0.000347 \\ 
55569.399842 & -792.5 & 0.000231 & 58384.376901 &    0.0 & 0.000115 & 58487.386712 &   29.0 & 0.000050 \\ 
55585.385154 & -788.0 & 0.000224 & 58386.149994 &    0.5 & 0.000036 & 58489.158941 &   29.5 & 0.000212 \\ 
55617.353910 & -779.0 & 0.000525 & 58387.929945 &    1.0 & 0.000032 & 58517.570308 &   37.5 & 0.000092 \\ 
55782.523263 & -732.5 & 0.006381 & 58389.703234 &    1.5 & 0.000034 & 58519.349322 &   38.0 & 0.000339 \\ 
55789.623554 & -730.5 & 0.004157 & 58391.483399 &    2.0 & 0.000029 & 58521.122208 &   38.5 & 0.000260 \\ 
55798.509614 & -728.0 & 0.006379 & 58393.256496 &    2.5 & 0.000038 & 58522.901643 &   39.0 & 0.000104 \\ 
55805.613433 & -726.0 & 0.000281 & 58395.036286 &    3.0 & 0.000045 & 58524.675031 &   39.5 & 0.000322 \\ 
55814.487262 & -723.5 & 0.001839 & 58396.808493 &    3.5 & 0.000041 & 58526.454370 &   40.0 & 0.000143 \\ 
55821.591741 & -721.5 & 0.000345 & 58398.588304 &    4.0 & 0.000043 & 58528.227732 &   40.5 & 0.000051 \\ 
55830.475747 & -719.0 & 0.000319 & 58400.360410 &    4.5 & 0.000038 & 58535.332344 &   42.5 & 0.000118 \\ 
55846.458539 & -714.5 & 0.000440 & 58402.139807 &    5.0 & 0.000043 & 58537.112508 &   43.0 & 0.000041 \\ 
55862.446155 & -710.0 & 0.000396 & 58403.912264 &    5.5 & 0.000038 & 58538.886252 &   43.5 & 0.000060 \\ 
55869.550282 & -708.0 & 0.000334 & 58405.691333 &    6.0 & 0.000036 & 58540.666174 &   44.0 & 0.000040 \\ 
55871.320641 & -707.5 & 0.000496 & 58407.463464 &    6.5 & 0.000178 & 58545.993103 &   45.5 & 0.000043 \\ 
55878.427212 & -705.5 & 0.000304 & 58409.242515 &    7.0 & 0.000204 & 58547.772304 &   46.0 & 0.000036 \\ 
55885.527682 & -703.5 & 0.001080 & 58411.014691 &    7.5 & 0.000062 & 58549.545373 &   46.5 & 0.000040 \\ 
55901.515712 & -699.0 & 0.000121 & 58412.794027 &    8.0 & 0.000039 & 58551.324561 &   47.0 & 0.000048 \\ 
55917.497684 & -694.5 & 0.000238 & 58414.566148 &    8.5 & 0.000041 & 58553.097355 &   47.5 & 0.000031 \\ 
55926.382119 & -692.0 & 0.000221 & 58416.345405 &    9.0 & 0.000037 & 58554.876377 &   48.0 & 0.000043 \\ 
55933.485404 & -690.0 & 0.000601 & 58418.117360 &    9.5 & 0.000053 & 58558.428144 &   49.0 & 0.000030 \\ 
55942.363865 & -687.5 & 0.000497 & 58421.669034 &   10.5 & 0.000058 & 58560.201140 &   49.5 & 0.000044 \\ 
55974.328517 & -678.5 & 0.000494 & 58423.448354 &   11.0 & 0.000059 & 58561.979498 &   50.0 & 0.000039 \\ 
56178.576926 & -621.0 & 0.000135 & 58425.219923 &   11.5 & 0.000044 & 58563.752076 &   50.5 & 0.000041 \\ 
56242.512478 & -603.0 & 0.000085 & 58426.999079 &   12.0 & 0.000040 & 58565.530780 &   51.0 & 0.000036 \\ 
56251.389203 & -600.5 & 0.000205 & 58428.771016 &   12.5 & 0.000044 & 58567.303422 &   51.5 & 0.000044 \\ 
56258.493033 & -598.5 & 0.000160 & 58430.550484 &   13.0 & 0.000043 & 58627.689806 &   68.5 & 0.000054 \\ 
56299.344637 & -587.0 & 0.000081 & 58432.322756 &   13.5 & 0.000038 & 58629.468525 &   69.0 & 0.000037 \\ 
56306.448300 & -585.0 & 0.000131 & 58434.101671 &   14.0 & 0.000036 & 58631.241022 &   69.5 & 0.000048 \\ 
56315.326180 & -582.5 & 0.000139 & 58435.874228 &   14.5 & 0.000044 & 58633.020688 &   70.0 & 0.000041 \\ 
56347.295223 & -573.5 & 0.000197 & 58439.425783 &   15.5 & 0.000151 & 58634.793046 &   70.5 & 0.000039 \\ 
56567.521093 & -511.5 & 0.000201 & 58441.205071 &   16.0 & 0.000106 & 58636.572329 &   71.0 & 0.000038 \\ 
56574.624240 & -509.5 & 0.001192 & 58442.977939 &   16.5 & 0.000106 & 58638.344386 &   71.5 & 0.000035 \\ 
56599.488418 & -502.5 & 0.000198 & 58444.757236 &   17.0 & 0.000055 & 58640.122770 &   72.0 & 0.000052 \\ 
56606.593197 & -500.5 & 0.000347 & 58446.530362 &   17.5 & 0.000081 & 58641.895837 &   72.5 & 0.000048 \\ 
56608.372742 & -500.0 & 0.000338 & 58448.310155 &   18.0 & 0.000159 & 58643.674949 &   73.0 & 0.000042 \\ 
56622.579359 & -496.0 & 0.000190 & 58451.862607 &   19.0 & 0.000129 & 58645.447054 &   73.5 & 0.000042 \\ 
56672.309088 & -482.0 & 0.000259 & 58453.635231 &   19.5 & 0.000199 & 58647.226031 &   74.0 & 0.000039 \\ 
56711.382703 & -471.0 & 0.000793 & 58455.414842 &   20.0 & 0.000183 & 58648.998180 &   74.5 & 0.000050 \\ 
58354.179002 &   -8.5 & 0.000053 & 58457.187794 &   20.5 & 0.000121 & 58650.777500 &   75.0 & 0.000032 \\ 
58355.957694 &   -8.0 & 0.000159 & 58458.967508 &   21.0 & 0.001603 & 58652.549690 &   75.5 & 0.000037 \\ 
58357.730220 &   -7.5 & 0.000091 & 58460.741232 &   21.5 & 0.000095 &              &        &          \\ 
58359.509098 &   -7.0 & 0.000090 & 58462.521455 &   22.0 & 0.000124 &              &        &          \\ 
58361.281906 &   -6.5 & 0.000122 & 58469.628209 &   24.0 & 0.000054 &		   &	    &	       \\ 
58363.061012 &   -6.0 & 0.000225 & 58471.401030 &   24.5 & 0.000064 &		   &	    &	       \\ 
58364.833939 &   -5.5 & 0.000113 & 58473.180483 &   25.0 & 0.000125 &		   &	    &	       \\ 
58370.165529 &   -4.0 & 0.000115 & 58474.952969 &   25.5 & 0.000072 &		   &	    &	       \\ 
\hline
\end{tabular}

{\em Notes.} Integer and half-integer cycle numbers refer to primary and secondary eclipses, respectively. Most of the eclipses in the first column (cycle nos. $-833.0$ to $-471.0$) were observed in the WASP project, while the newer eclipse times (from cycle no. $-8.5$) determined from the {\em TESS} measurements.
\end{table*}

\subsection{SED data and Gaia results}
\label{Sect:SEDGaia}

Despite the relative brightnesses of both EBs we have found only a very limited number of spectroscopic measurements in the literature (without any indications of the multiplicity of the sources).  In particular, we found no RV data during our literature searches.  As a consequence, no dynamically constrained masses are available for these systems. Furthermore, although the spectroscopic survey {\em TESS}-HERMES DR-1 \citep{TESSHERMES} gives spectroscopically determined effective temperatures for both systems, the surface gravities ($\log g$) derived from the same spectra clearly contradict our lightcurve solutions.  Therefore, we utilized a combination of SED data, PARSEC theoretical stellar isochrones \citep{PARSEC} and photodynamical model solutions (see Sect.\,\ref{sec:dyn}, below) to determine the stellar masses and temperatures. In order to do this, we took compiled $J$, $H$, $K_s$, $W_1$, and $W_2$ magnitudes from the eighth version of the {\em TESS} Input Catalog \citep[TIC-8,][]{TIC}, which in turn subsumes photometric data from a large number of other photometric catalogs such as the Two-Micron All-sky Survey \citep[2MASS,][]{2MASS}, Wide-field Infrared Survey Explorer \citep[WISE,][]{WISE}, etc. Moreover, we collected Johnson $B$, $V$ and Sloan $g'$, $r'$, $i'$ magnitudes from the ninth data release of the AAVSO Photometric All-Sky Survey  \citep[APASS9,][]{APASS} and also $B_T$, $V_T$ magnitudes from Tycho-2 catalog \citep{TYCHO2}. Furthermore, we also utilized Gaia $G$, $B_P$ and $R_P$ magnitudes, and trigonometric parallax $\varpi_\mathrm{DR2}$ taken from Gaia DR2 \citep{GaiaDR2}. Finally, we also took from TIC-8 the metallicity $[M/H]$ obtained from the spectroscopic survey {\em TESS}-HERMES DR-1 \citep{TESSHERMES}.

We also consulted the Gaia DR2 and the Hipparcos/Tycho data bases for varying proper motion results. Both targets are included in the Tycho-2 catalog \citep{TYCHO2}. There are no significant proper motion changes during the 24.25 years long baseline for our targets. For TIC\,220397947 the proper motion errors are very close to the median value of the corresponding brightness range (see Table 2. of \citealp{TYCHO2}) and are consistent with the Gaia data within 2\,$\sigma$. The errors for TIC\,167692429 are slightly larger, which can be due to the smaller number of observations. In order to be able to include proper motion data into the analysis we have to wait for the Gaia final data release which will contain all epoch and transit observations. All data listed above are presented in Table\,\ref{tab:object}.

\section{Joint Physical and Dynamical modelling of all the available observational data}
\label{sec:dyn}

In our previous work \citep{Borkovits2019a,Borkovits2019b} we carried out joint, simultaneous spectro-photodynamical analyses of light-, ETV-, and radial-velocity (RV) curves of a number of compact hierarchical multiple systems. In those cases there were available one or more RV curves and, therefore, we were able to determine model-independent, dynamical masses for each component.\footnote{For a dynamically interacting system one needs only one component's mass (i.e. one RV amplitude), at least in theory, as the short-term dynamical interactions constrain the mass ratios strongly \citep[see, e.\,g.][]{rappaportetal13,Borkovits2015}.}

In the present situation, however, in the absence of RV measurements we adopted an alternative, and astrophysical model-dependent method, inferring stellar masses and temperatures from the combined modelling of lightcurves, ETVs, multiple SEDs, and stellar evolution models. 

The combination of stellar isochrones and/or SED fits with an eclipsing binary lightcurve solver was introduced previously by, e.g.,  \citet{devorcharbonneau06} who pointed out that this method could lead to reasonable mass estimations for a large number of faint EBs observed during large photometric surveys. Later, \citet{moedistefano13,moedistefano15} analysed hundreds of EBs in the LMC in a similar manner. A related empirical method has also been used by \citet{maxtedhutcheon18} for characterizing EBs from the {\em K2} survey.  Most recently, \citet{windemuthetal19} have determined physical and orbital parameters in such a manner for the detached EBs in the original {\em Kepler} field. Our method was mainly inspired by the paper of \citet{windemuthetal19}, however, as far as we are aware, our efforts are the first to apply the SED+isochrone fitting method for multiple stellar systems.

For the combined modeling, we incorporated into the software package {\sc Lightcurvefactory} \citep[see][and further references therein]{Borkovits2019a,Borkovits2019b} the ability to handle tables of stellar isochrones and also to fit isochrone generated SED data to the observed magnitudes making use of the known Gaia distance. 

For this purpose we generated machine readable PARSEC stellar isochrone grids \citep{PARSEC} via the web based tool CMD 3.3\footnote{\url{http://stev.oapd.inaf.it/cgi-bin/cmd}}. The table(s) contain initial and actual stellar masses, bolometric luminosities ($\log L/L_\odot$), effective temperatures ($\log T_\mathrm{eff}$), surface gravities ($\log g$), as well as absolute stellar passband magnitudes in several photometric systems for the user selected grid of stellar metallicities and ages ($\log\tau$). {\sc Lightcurvefactory} now uses this table to calculate the above listed parameters (with trilinear interpolation) for the set of (mass, metallicity, age) values of the given star(s) under analysis. Then for the lightcurve analysis part of the problem, the obtained effective temperatures and stellar radii can be used directly to generate the model lightcurve, while for the SED fitting the interpolated absolute passband magnitudes are converted to observed ones, taking into account the interstellar extinction and the distance of the system.  To calculate the interstellar reddening, following the treatment of TIC-8 catalog, we corrected for line-of-sight dust extinction assuming a standard exponential model for the dust with a scale height of 125\,pc. 

The main steps of our joint analysis were as follows.
\begin{itemize}
\item[(i)] First, we carried out a joint photodynamical lightcurve and ETV analysis of both systems, applying a Markov Chain Monte Carlo (MCMC) parameter search. Initially, we modelled both systems as hierarchical triple stars. 

Regarding the {\em TESS} SC lightcurves, in order to reduce the computational time we binned the two-minute cadence data, averaging them every half hour (i.e., 1800 s). Then, we narrowed the lightcurves to be modelled to $\pm0\fp03$ and $\pm0\fp04$ phase-domain regions around each eclipses, for TICs\,167692429 and 220397947, respectively. Then these lightcurves were used in the photodynamical analysis (applying, of course an appropriate cadence time correction).

Having obtained the WASP observations of TIC\,220397947, we realized that systematic departures from the expected (i.e., back-projected) eclipse times occurred one decade before the {\em TESS} observations, as well as a different eclipse period for the inner binary at that time. This was evident not only from an extended ETV analysis, but also directly from the fitted WASP lightcurves which contain several only partially observed eclipses (and were not included into the ETV analysis). Therefore, we decided to model TIC\,220397947 as a quadruple stellar system, having a 2+1+1 hierarchy. Therefore, besides the {\em TESS} SC observations, we included in the MCMC search another lightcurve file, containing the similarly narrow sections around all the eclipses observed (mostly partially) with the WASP cameras. Furthermore, the times of minima deduced from the WASP observations were also added to the ETV curves to be fitted.

Similar to our previous work, in these runs we adjusted the following parameters:
\begin{itemize}
\item[--]{Three parameters related to the orbital elements of the inner binaries. For TIC\,167692429 these parameters were as follows: the eccentricity ($e_1$), the phase of the secondary eclipse relative to the primary one ($\phi_\mathrm{sec,1}$) which constrains the argument of periastron ($\omega_1$, see \citealt{Rappaport2017}), and the inclination ($i_1$). For TIC\,220397947 however, because of the very small eccentricity of the eclipsing pair, adjusting the more commonly used parameters $(e\sin\omega)_1$ and $(e\cos\omega)_1$ was found to be more practical.\footnote{Two other inner-orbit related parameters, namely the instantaneous orbital periods ($P_1$) and inferior conjunction time $(\mathcal{T}_0)_1$ of the secondary components of the inner binaries, i.e., the mid-primary-eclipse-times, were constrained with the use of the ETV curves in the manner explained in Appendix\,A of \citet{Borkovits2019a}, while the sixth orbital element, $\Omega_1$, as irrelevant, was kept fixed at zero.}}
\item[--]{Six parameters related to the orbital elements of the wide orbit of the third component: $P_2$, $(e\sin\omega)_2$, $(e\cos\omega)_2$, $i_2$, the time of periastron passage of star C along its wide orbit ($\tau_2$), and the position angle of the node of the wide orbit ($\Omega_2$).\footnote{As $\Omega_1=0\degr$ was assumed at epoch $t_0$ for all runs, $\Omega_2$ set the initial trial value of the differences of the nodes (i.~e., $\Delta\Omega$), which is the truly relevant parameter for dynamical modelling.} Furthermore, in the case of the analysis runs for 2+1+1 quadruple representing TIC\,220397947 a similar set of the orbital parameters ($P_3$, $[e\sin\omega]_3$, $[e\cos\omega]_3$, $i_3$, $\tau_3$ and, $\Omega_3$) were adjusted for the outermost orbit.}
\item[--]{Two (or three) mass-related parameters: the mass ratios of the two (or three) orbits $q_1$, $q_2$, (and $q_3$);}
\item[--]{and, finally, four (five) other parameters which are related (almost) exclusively to the lightcurve solutions, as follows: the duration of the primary eclipse $(\Delta t)_\mathrm{pri}$ closest to epoch $t_0$ (which is an observable that is strongly connected to the sum of the fractional radii of stars A and B, i.e., scaled by the inner semi-major axis, see \citealt{Rappaport2017}), the ratio of the radii of stars A and B ($R_\mathrm{B}/R_\mathrm{A}$), and the temperature ratios of $T_\mathrm{B}/T_\mathrm{A}$, and the passband-dependent extra light(s) $\ell_\mathrm{TESS}$ (and $\ell_\mathrm{WASP}$).} 
\end{itemize} 

 Turning to the other, lightcurve-related parameters, we applied a logarithmic limb-darkening law, where the coefficients were interpolated during each trial step from the pre-computed passband-dependent tables in the {\sc Phoebe} software \citep{Phoebe}. The {\sc Phoebe}-based tables, in turn, were derived from the stellar atmospheric models of \citet{castellikurucz04}. Due to the nearly spherical stellar shapes in both inner binaries, accurate settings of gravity darkening coefficients have no influence on the lightcurve solution and, therefore, we simply adopted a fixed value of $g=0.32$ which is appropriate for stars having a convective envelope according to the traditional model of \citet{lucy67}.  Regarding the illumination/reradiation effect we found that it was quite negligible for the eclipsing pair of TIC\,167692429, while it had a minor effect $(\lesssim500\,$ppm) for the lightcurve of TIC\,220397947. Therefore, in order to save computing time, this effect was neglected. On the other hand, the Doppler-boosting effect \citep{loebgaudi03,vankerkwijketal10} which was also found to be negligible, but needs only very minor additional computational costs, was included into our model.

Moreover, in this first stage of analysis we set the unadjusted primary masses ($m_\mathrm{A}$) and effective temperatures ($T_\mathrm{A}$) to the values given in the TIC. Here we emphasize, that at this stage the actual value of the masses and temperatures of the primaries played only a minor role, since we used these runs largely to constrain  temperature ratios of the EBs, as well as the mass ratios of both the inner and outer binaries. 

These runs revealed that both inner binaries were comprised of similar stars (i.e., both the inner mass and temperature ratios were found to be close to unity). Furthermore, from these lightcurve solutions we were able to obtain reasonable estimates for the local surface gravities of each EB member star. For TIC\,167692429 we found $\log g\approx4.0-4.1$ for both stars, while for TIC\,220397947 it was found to be $\log g\approx4.1-4.3$ and $\approx4.2-4.4$ for the primary and the secondary, respectively. These sets of the preliminary solutions have also shown that the additional outer stellar components are less massive stars which add only minor contributions to the systems' brighnesses and, therefore, can safely be omitted for the next step, i.e. for the preliminary SED fitting of the EBs.

\item[(ii)] In the next step we fitted the observed passband magnitudes\footnote{Note, for all of the SED-fitting processes we arbitrarily multiplied the small uncertainties of Gaia's $G$, $G_{BP}$, $G_{RP}$ magnitudes by a factor of ten for two reasons. First, we wanted to avoid the extreme over-dominance of these three magnitudes during the $\chi^2$-optimization processes.  The second reason was to counterbalance the expected larger systematic errors in the model SED magnitudes that were interpolated from the grid points. The uncertainties in the Gaia magnitudes are two orders of magnitude smaller than for the other SED points as well as compared to the systematic errors in the model SED.  We therefore decided to adopt an uncertainty for the Gaia magnitudes that is the geometric mean between the actual Gaia magnitude uncertainties and the other uncertainties in this part of the problem.  Substantially smaller uncertainties for the Gaia points would render the other SED points of little value, while larger uncertainties for the Gaia magnitudes would fail to make use of their high precision.} (see Table\,\ref{tab:object}) to SED models to find the approximative temperatures of the binary members. We fixed the inner mass ratios ($q_1$) to the values obtained previously in step (i), while a preliminary value for the primaries' masses were again taken from the TIC catalogue. With these stellar masses we initialized an SED fitting procedure with the use of the built-in MCMC solver of our code. At this stage we adjusted only the stellar age ($\log\tau$) parameter, while stellar metallicities and the interstellar extinction parameter $E(B-V)$ were kept fixed at their catalog values. Moreover, the photometric distance was recalculated at each trial step so as to minimize $\chi^2_\mathrm{SED}$. In such a manner we quickly found realistic $T_\mathrm{eff}$ values for the binary member's temperatures. 

\item[(iii)] Then, using these temperatures, $\log g_\mathrm{A}$\footnote{Strictly speaking, in the absence of a priori knowledge of the primary's mass, the photodynamical lightcurve analysis results in only $\log g_\mathrm{A}*=\log g_\mathrm{A}-\frac{1}{3}\log m_\mathrm{A}$, \citep[see, e.g.][]{hajduetal17} from which value, however, $\log g_\mathrm{A}$ can be easily calculated for each individual trial mass. Then, the trivial relation of $\log g_\mathrm{B}=\log g_\mathrm{A}+\log q_1-2\log(r_\mathrm{B}/r_\mathrm{A})$ gives the surface gravity of the secondary.}, mass ratio ($q_1$), and the ratio of the radii ($r_\mathrm{B}/r_\mathrm{A}$) obtained in the previous lightcurve fits, we searched the interpolated\footnote{During this search process the primary's actual mass ($m_\mathrm{A}$) was the only free parameter, which was increased evenly between 0.5\,M$_\odot$ and 3.0\,M$_\odot$ with a stepsize of 0.01\,M$_\odot$. At the same time, we calculated the secondary's mass from the mass ratio ($q_1$). Then for all the numerous doublets (metallicity, log age) in the PARSEC table, we calculated the other astrophysical parameters (i.e. $T_\mathrm{eff}$-s, $\log g$-s, and $R$-s) for both masses ($m_\mathrm{A}$ and $m_\mathrm{B}$), linearly interpolating between the previous and the next mass grid elements.} PARSEC isochrone grids for those items (i.e. age, metallicity, mass triplets) where both the primary's and secondary's $\log T_\mathrm{eff}$-s and $\log g$-s were simultaneously within a few percent of the values obtained in the previous steps.  These grid items (i.e. metallicity, age, and primary stellar mass values) were selected as the initial trial values for the first round of the combined SED (isochrone) and lightcurve fits.  Because we understood from step (i) that the outer stellar components yield only a minor contribution to the SED, in order to save substantial computational time at this point, we formed and fit folded lightcurves from the {\em TESS} SC observations of the two binaries. (Note, in the case of TIC\,167692429 we used only the Sector 1--9 data, when the EB's lightcurve showed only minor variations in both eclipse shape and timing.) Then we binned these curves into 2000 uniform phase cells around the two eclipses and 500 cells in the out of eclipse sections. For each cell we kept only the average of the individual flux values within the given cell.

These phased lightcurves were fitted simultaneously with the SED, using the corresponding, interpolated PARSEC isochrones. The initial values of the primary mass ($m_1$), stellar age ($\log\tau$) and metallicity ($[M/H]$) were taken from the above mentioned parameter triplets. During the MCMC runs these variables were adjusted together with the mass ratio ($q_1$), the orbital element-related parameters (see above), the third light ($\ell$), the extinction parameter [E(B-V)] and the distance of the actual system ($d$). Regarding this last item, our treatment slightly departs from that of \citet{windemuthetal19}. Those authors used Gaussian priors for the distance calculated from Gaia's DR2 trigonometric parallax (and its uncertainty). We, however, used a uniform prior, initializing the distance variable with the Gaia trigonometric distance, but allowing practically any distance and, therefore, not penalizing any departures from the Gaia DR2 distances.  The use of a uniform prior instead of a Gaia parallax-based Gaussian prior can be justified because at this stage we have not yet considered the other stellar components of these multiple star systems.  And, since a minor part of the targets' total fluxes was thereby omitted, the binaries would be slightly fainter and, consequently, seemingly closer than the complete multiple systems.

As a conclusion for this stage of the analysis we obtained dozens of lightcurve solutions for a large variety of stellar age, metallicity and mass triplets which were equally satisfactory or, quantitatively, where we found that $\chi^2\lesssim1.1\times\chi^2_\mathrm{min}$.

\item[(iv)] For the final stage in the analysis, our original intention was to select those solutions from the previous stage where the inferred photometric distance was within the 3-$\sigma$ uncertainty of the Gaia distance, and initialize the joint photodynamical lightcurve, ETV curve, SED and PARSEC isochrone analysis with these parameters. However, we found that for TIC\,167692429 all former stage solutions resulted in an incompatible distance with the Gaia result. Therefore, for this triple star, instead of the distances, we have chosen those solutions which were compatible with the metallicity given in the TIC. Oppositely, for TIC\,220397947 we used the distances for the selection. 

At this stage the radii and effective temperatures of all three (four) stars  were constrained by PARSEC isochrones. Apart from these, the adjusted parameters were the same as those listed in item (i) and item (iii) above. Moreover, in the case of TIC\,220397947 we applied a Gaussian prior to the distance peaked at Gaia's result; however we set $\sigma=3\sigma_\mathrm{Gaia}$, allowing for the inclusion of some systematic effects that might be present in the Gaia DR2 results due to the multiplicity of stars. On the other hand, for TIC\,167692429 we kept a uniform prior on the distance. 

A flow diagram of the entire fitting process is drawn in Fig.\,\ref{fig:flow}. 
\end{itemize}

\begin{figure*}
\centerline{
\includegraphics[width=0.98\textwidth]{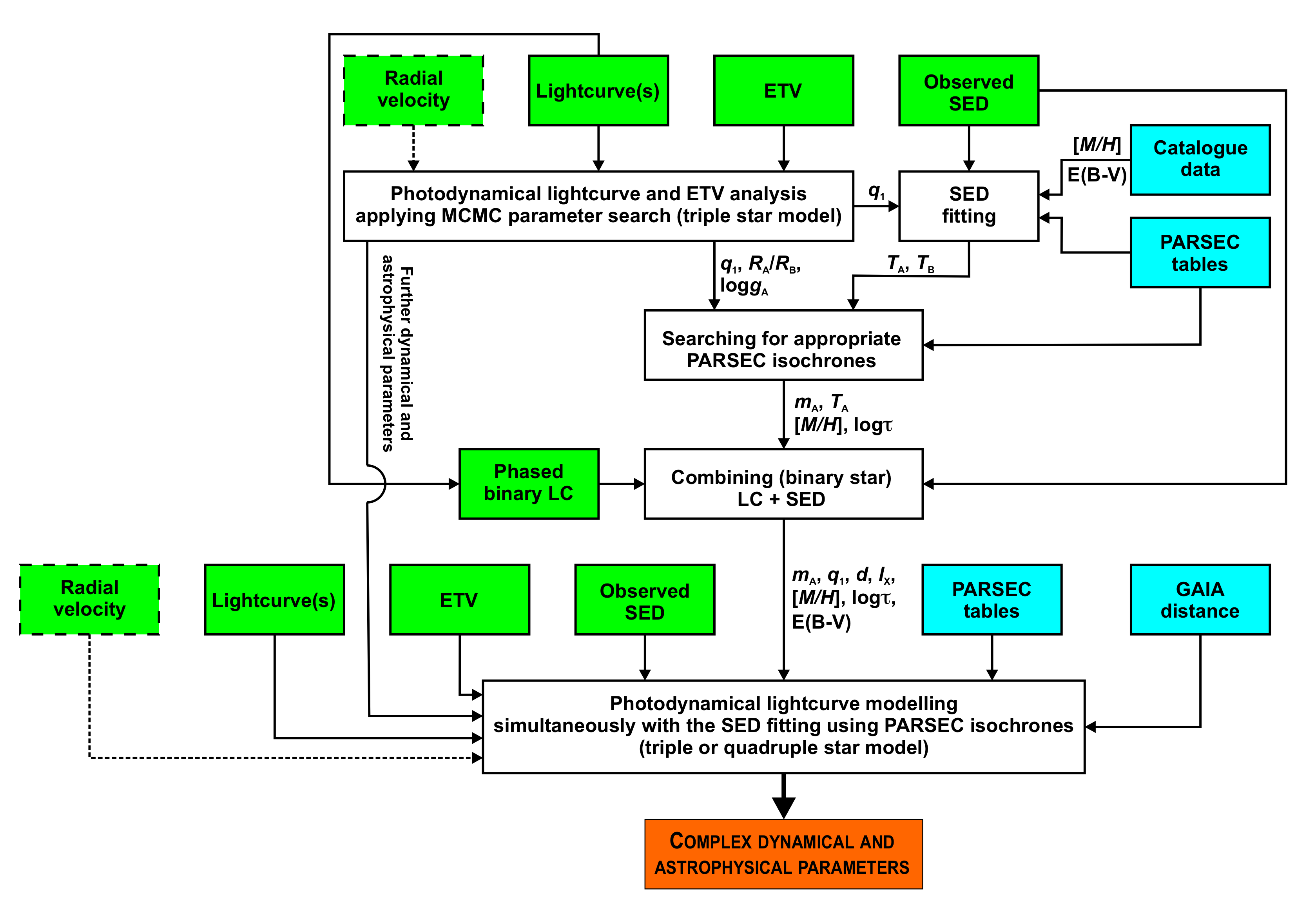}
}
\caption{Flow diagram for the entire combined fitting analysis. We also list radial velocity data in the rectangle with dashed borders. RV data were not used, however, in the present analysis because they were unavailable for these two systems; but, in principle, these can also be used in the combined analysis.}
\label{fig:flow}
\end{figure*}
 
The orbital and astrophysical parameters derived from the photodynamical analysis are tabulated in Tables\,\ref{tab: syntheticfit_TIC167} and \ref{tab: syntheticfit_TIC220}, and will be discussed in the subsequent Sections\,\ref{sec:par} and \ref{sec:orbprop}. The corresponding model lightcurves are presented in Figs.\,\ref{fig:eclipsefitT167}, \ref{fig:eclipsefitT220}, while the model ETV curves plotted against the observed ETVs is shown in Figs.\,\ref{fig:ETV}, \ref{fig:ETV2}.

\begin{table*}
 \centering
 \caption{Orbital and astrophysical parameters of TIC\,167692429 from the joint photodynamical lightcurve, ETV, SED and PARSEC isochrone solution. Besides the usual observational system of reference related angular orbital elements ($\omega$, $i$, $\Omega$), their counterparts in the system's invariable plane related dynamical frame of reference are also given ($\omega^\mathrm{dyn}$, $i^\mathrm{dyn}$, $\Omega^\mathrm{dyn}$). Moreover, $i_\mathrm{m}$ denotes the mutual inclination of the two orbital planes, while $i_\mathrm{inv}$ and $\Omega_\mathrm{inv}$ give the position of the invariable plane with respect to the tangential plane of the sky (i.\,e., in the observational frame of reference). Columns 2--4 represent post-MS solutions, while columns 5--7 list the results of pre-MS solutions.}
 \label{tab: syntheticfit_TIC167}
\begin{tabular}{@{}lllllll}
  \hline
&\multicolumn{3}{c}{Post-MS solution}&\multicolumn{3}{c}{Pre-MS solution}\\
\hline
\multicolumn{7}{c}{orbital elements$^a$} \\
\hline
   & \multicolumn{3}{c}{subsystem} & \multicolumn{3}{c}{subsystem} \\
   & \multicolumn{2}{c}{A--B} & AB--C  & \multicolumn{2}{c}{A--B} & AB--C\\
  \hline
  $P$ [days] & \multicolumn{2}{c}{$10.26276\pm0.00012$} & $331.50_{-0.33}^{+0.28}$ & \multicolumn{2}{c}{$10.26286_{-0.00011}^{+0.00010}$} & $331.45\pm0.31$  \\
  $a$ [R$_\odot$] & \multicolumn{2}{c}{$27.25_{-0.41}^{+0.43}$} & $304.6_{-4.9}^{+4.5}$ & \multicolumn{2}{c}{$28.00_{-0.24}^{+0.20}$} & $312.9_{-2.3}^{+2.5}$\\
  $e$ & \multicolumn{2}{c}{$0.1734_{-0.0008}^{+0.0009}$} & $0.55749_{-0.00071}^{+0.00072}$ & \multicolumn{2}{c}{$0.1723\pm0.0010$} & $0.55730_{-0.00082}^{+0.00081}$\\
  $\omega$ [deg]& \multicolumn{2}{c}{$288.35_\pm0.11$} & $0.89_{-0.17}^{+0.19}$ & \multicolumn{2}{c}{$288.47\pm0.14$} & $0.79_{-0.21}^{+0.17}$\\ 
  $i$ [deg] & \multicolumn{2}{c}{$85.731_{-0.039}^{+0.036}$} & $85.68_{-0.23}^{+0.22}$ & \multicolumn{2}{c}{$85.796_{-0.051}^{+0.057}$} & $85.50_{-0.25}^{+0.34}$\\
  $\tau$ [BJD - 2400000]& \multicolumn{2}{c}{$58320.6874_{-0.0031}^{+0.0030}$} & $58611.987\pm0.078$ & \multicolumn{2}{c}{$58320.6908_{-0.0038}^{+0.0039}$} & $58611.902_{-0.109}^{+0.099}$\\
  $\Omega$ [deg] & \multicolumn{2}{c}{$0.0$} & $-27.26_{-0.13}^{+0.14}$ & \multicolumn{2}{c}{$0.0$} & $-27.43\pm0.16$\\
  $i_\mathrm{m}$ [deg] & \multicolumn{3}{c}{$27.18_{-0.14}^{+0.13}$} & \multicolumn{3}{c}{$27.36\pm0.16$}\\
  $\omega^\mathrm{dyn}$ [deg]& \multicolumn{2}{c}{$199.52_{-0.53}^{+0.50}$} & $89.98_{-0.47}^{+0.46}$ & \multicolumn{2}{c}{$200.14_{-0.75}^{+0.48}$} & $89.88_{-0.63}^{+0.57}$\\
  $i^\mathrm{dyn}$ [deg] & \multicolumn{2}{c}{$20.93_{-0.14}^{+0.13}$} & $6.246_{-0.018}^{+0.019}$& \multicolumn{2}{c}{$21.12\pm0.16$} & $6.232\pm0.020$\\
  $\Omega^\mathrm{dyn}$ [deg] & \multicolumn{2}{c}{$270.44_{-0.47}^{+0.48}$} & $90.44_{-0.47}^{+0.48}$ & \multicolumn{2}{c}{$269.98_{-0.47}^{+0.66}$} & $89.98_{-0.47}^{+0.66}$\\
  $i_\mathrm{inv}$ [deg] & \multicolumn{3}{c}{$85.60_{-0.18}^{+0.17}$} & \multicolumn{3}{c}{$85.48_{-0.21}^{+0.26}$}\\
  $\Omega_\mathrm{inv}$ [deg] & \multicolumn{3}{c}{$-20.99_{-0.13}^{+0.14}$} & \multicolumn{3}{c}{$-21.18\pm0.16$}\\
  \hline
  mass ratio $[q=m_\mathrm{sec}/m_\mathrm{pri}]$ & \multicolumn{2}{c}{$1.005_{-0.030}^{+0.011}$} & $0.337\pm0.002$ & \multicolumn{2}{c}{$0.993\pm0.003$} & $0.341\pm0.003$\\
  $K_\mathrm{pri}$ [km\,s$^{-1}$] & \multicolumn{2}{c}{$68.46_{-1.90}^{+0.99}$} & $14.07_{-0.24}^{+0.23}$ & \multicolumn{2}{c}{$69.63_{-0.67}^{+0.53}$} & $14.58\pm0.14$ \\ 
  $K_\mathrm{sec}$ [km\,s$^{-1}$] & \multicolumn{2}{c}{$68.04\pm0.65$} & $41.78_{-0.61}^{+0.64}$ & \multicolumn{2}{c}{$70.16_{-0.57}^{+0.43}$} & $42.81_{-0.37}^{+0.26}$\\ 
  \hline  
\multicolumn{7}{c}{stellar parameters} \\
\hline
   & A & B &  C & A & B &  C \\
  \hline
 \multicolumn{7}{c}{Relative quantities$^b$} \\
  \hline
 fractional radius [$R/a$]  & $0.0657_{-0.0016}^{+0.0034}$ & $0.0667_{-0.0040}^{+0.0013}$  & $0.00263_{-0.00009}^{+0.00008}$ & $0.0663\pm0.0003$ & $0.0660\pm0.0004$  & $0.00350_{-0.00002}^{+0.00003}$\\
 fractional flux [in {\em TESS}-band] & $0.5316$  & $0.3999$    & $0.0576$ & $0.4788$  & $0.4567$    & $0.0497$\\
 fractional flux [in {\em SWASP}-band]& $0.5370$  & $0.4170$    & $0.0460$ & $0.4873$  & $0.4813$    & $0.0314$\\
 \hline
 \multicolumn{7}{c}{Physical Quantities} \\
  \hline 
 $m$ [M$_\odot$] & $1.284_{-0.043}^{+0.056}$ & $1.295_{-0.076}^{+0.059}$ & $0.869_{-0.043}^{+0.039}$ & $1.402_{-0.035}^{+0.028}$ & $1.390_{-0.037}^{+0.030}$ & $0.949_{-0.021}^{+0.026}$\\
 $R^b$ [R$_\odot$] & $1.800_{-0.046}^{+0.075}$ & $1.822_{-0.131}^{+0.043}$ & $0.803_{-0.040}^{+0.036}$ & $1.854\pm0.020$ & $1.847_{-0.026}^{+0.022}$ & $1.094\pm0.007$\\
 $T_\mathrm{eff}^b$ [K]& $6544_{-27}^{+20}$ & $6518_{-34}^{+23}$ & $5597_{-112}^{+34}$ & $6612_{-28}^{+31}$ & $6531_{-28}^{+26}$ & $4632_{-41}^{+49}$\\
 $L_\mathrm{bol}^b$ [L$_\odot$] & $5.33_{-0.35}^{+0.49}$ & $5.32_{-0.74}^{+0.32}$ & $0.57_{-0.10}^{+0.06}$ & $5.90_{-0.13}^{+0.12}$ & $5.574_{-0.20}^{+0.16}$ & $0.49\pm0.02$\\
 $M_\mathrm{bol}^b$ & $2.95_{-0.10}^{+0.08}$ & $2.95_{-0.06}^{+0.16}$ & $5.38_{-0.10}^{+0.20}$ & $2.84\pm0.02$ & $2.90_{-0.03}^{+0.04}$ & $5.53_{-0.05}^{+0.04}$\\
 $M_V^b           $ & $2.95_{-0.09}^{+0.08}$ & $2.96_{-0.07}^{+0.17}$ & $5.48_{-0.11}^{+0.22}$ & $2.84\pm0.03$ & $2.91_{-0.04}^{+0.05}$ & $6.02_{-0.09}^{+0.08}$\\
 $\log g^b$ [dex] & $4.04_{-0.05}^{+0.02}$ & $4.03_{-0.01}^{+0.04}$ & $4.57_{-0.02}^{+0.02}$ & $4.046\pm0.004$ & $4.047\pm0.004$ & $4.336\pm0.008$\\
 \hline
$\log$(age) [dex] &\multicolumn{3}{c}{$9.464_{-0.052}^{+0.056}$} &\multicolumn{3}{c}{$7.028_{-0.032}^{+0.030}$}\\
$[M/H]$  [dex]    &\multicolumn{3}{c}{$-0.220_{-0.059}^{+0.112}$} &\multicolumn{3}{c}{$-0.207_{-0.128}^{+0.103}$}\\
$E(B-V)$ [mag]    &\multicolumn{3}{c}{$0.0533_{-0.0032}^{+0.0036}$} &\multicolumn{3}{c}{$0.0499_{-0.0049}^{+0.0033}$}\\
extra light $\ell_4$  [in {\em TESS}-band]&\multicolumn{3}{c}{$0.079_{-0.012}^{+0.010}$} &\multicolumn{3}{c}{$0.114_{-0.013}^{+0.012}$}\\
extra light $\ell_4$  [in {\em SWASP}-band]&\multicolumn{3}{c}{$0.0$} &\multicolumn{3}{c}{$0.0$}\\
$(M_V)_\mathrm{tot}^b$           &\multicolumn{3}{c}{$2.17_{-0.07}^{+0.03}$} &\multicolumn{3}{c}{$2.09\pm0.03$}\\
distance [pc]                &\multicolumn{3}{c}{$552_{-7}^{+14}$}  &\multicolumn{3}{c}{$574\pm8$}\\  
\hline
\end{tabular}

{\em Notes. }{$a$: Instantaneous, osculating orbital elements, calculated for epoch $t_0=2458310.0000$ (BJD); $b$: Interpolated from the PARSEC isochrones; }
\end{table*}

\begin{table*}
 \centering
 \caption{Orbital and astrophysical parameters of TIC\,220397947 from the joint photodynamical lightcurve, ETV, SED and PARSEC isochrone solution. Besides the usual observational system of reference related angular orbital elements ($\omega$, $i$, $\Omega$) their counterparts in the system's invariable plane related, dynamical frame of reference are also given ($\omega^\mathrm{dyn}$, $i^\mathrm{dyn}$, $\Omega^\mathrm{dyn}$). Moreover, $(i_\mathrm{m})_\mathrm{X-Y}$ denote the mutual inclination angle between orbits of subsystems X and Y, while $i_\mathrm{inv}$ and $\Omega_\mathrm{inv}$ gives the position of the invariable plane with respect to the tangential plane of the sky (i.e., in the observational frame of reference). }
 \label{tab: syntheticfit_TIC220}
\begin{tabular}{@{}lllll}
  \hline
\multicolumn{5}{c}{orbital elements$^a$} \\
\hline
   & \multicolumn{4}{c}{subsystem} \\
   & \multicolumn{2}{c}{A--B} & AB--C & ABC--D \\
  \hline
  $P$ [days] & \multicolumn{2}{c}{$3.55106_{-0.00005}^{+0.00004}$} & $77.083_{-0.010}^{+0.012}$ & $2661_{-29}^{+31}$  \\
  $a$ [R$_\odot$] & \multicolumn{2}{c}{$12.836_{-0.021}^{+0.037}$} & $107.544_{-0.951}^{+0.958}$ & $1167.0_{-8.9}^{+25.5}$ \\
  $e$ & \multicolumn{2}{c}{$0.00105_{-0.00034}^{+0.00041}$} & $0.2252_{-0.0124}^{+0.0195}$ & $0.526_{-0.006}^{+0.003}$ \\
  $\omega$ [deg]& \multicolumn{2}{c}{$319.441_{-15.38}^{+13.90}$} & $340.90_{-0.67}^{+0.43}$ & $198.10_{-0.51}^{+0.50}$ \\ 
  $i$ [deg] & \multicolumn{2}{c}{$82.283_{-0.182}^{+0.266}$} & $82.669_{-0.181}^{+0.274}$ & $88.624_{-3.9}^{+2.4}$\\
  $\tau$ [BJD - 2400000]&\multicolumn{2}{c}{$55411.8286_{-0.1353}^{+0.1559}$} & $55378.648_{-0.045}^{+0.037}$ & $53851.6_{-32.0}^{+32.5}$ \\
  $\Omega$ [deg] & \multicolumn{2}{c}{$0.0$} & $0.425_{-0.263}^{+0.177}$ & $17.9_{-3.8}^{+3.6}$ \\
  $(i_\mathrm{m})_\mathrm{AB-C,D}$ [deg] & \multicolumn{2}{c}{$-$} & $0.57_{-0.14}^{+0.24}$ & $18.9_{-4.2}^{+3.3}$\\
  $(i_\mathrm{m})_\mathrm{ABC-D}$ [deg]   & \multicolumn{2}{c}{$-$} & $-$ & $18.4_{-3.9}^{+3.3}$\\
  $\omega^\mathrm{dyn}$ [deg]& \multicolumn{2}{c}{$68.7_{-13.9}^{+17.0}$} & $88.8_{-13.5}^{+9.2}$ &$127.7_{-12.3}^{+8.7}$\\
  $i^\mathrm{dyn}$ [deg] & \multicolumn{2}{c}{$9.36_{-2.76}^{+1.84}$} & $8.84_{-2.45}^{+1.96}$ &$9.53_{-1.60}^{+1.68}$\\
  $\Omega^\mathrm{dyn}$ [deg] & \multicolumn{2}{c}{$69.8_{-8.3}^{+12.4}$} & $71.3_{-9.1}^{+13.0}$ & $250.8_{-8.9}^{+12.7}$\\
  $i_\mathrm{inv}$ [deg] & \multicolumn{4}{c}{$85.45_{-2.26}^{+1.42}$} \\
  $\Omega_\mathrm{inv}$ [deg] & \multicolumn{4}{c}{$8.86_{-2.72}^{+2.02}$} \\
  \hline
  mass ratio $[q=m_\mathrm{sec}/m_\mathrm{pri}]$ & \multicolumn{2}{c}{$0.950_{-0.012}^{+0.008}$} & $0.248_{-0.028}^{+0.023}$ & $0.081_{-0.012}^{+0.016}$ \\
  $K_\mathrm{pri}$ [km\,s$^{-1}$] & \multicolumn{2}{c}{$88.328_{-0.477}^{+0.318}$} & $14.296_{-1.403}^{+1.140}$ & $1.959_{-0.311}^{+0.369}$ \\ 
  $K_\mathrm{sec}$ [km\,s$^{-1}$] & \multicolumn{2}{c}{$92.941_{-0.550}^{+0.804}$} & $57.589_{-0.752}^{+1.043}$ & $24.188_{-0.251}^{+0.344}$ \\ 
  \hline  
\multicolumn{5}{c}{stellar parameters} \\
\hline
   & A & B &  C & D\\
  \hline
 \multicolumn{5}{c}{Relative quantities} \\
  \hline
 fractional radius [$R/a$]  & $0.0942_{-0.0012}^{+0.0010}$ & $0.0944_{-0.0012}^{+0.0012}$  & $0.00746_{-0.00031}^{+0.00025} $ & $0.00047_{-0.00003}^{+0.00003}$\\
 fractional flux [in {\em TESS}-band]& $0.5041$  & $0.4643$    & $0.0244$ & $0.0044$ \\
 fractional flux [in SWASP-band]     & $0.5233$  & $0.4648$    & $0.0101$ & $0.0017$ \\
 \hline
 \multicolumn{5}{c}{Physical Quantities$^b$} \\
  \hline 
 $m$ [M$_\odot$] & $1.152_{-0.009}^{+0.018}$ & $1.095_{-0.006}^{+0.005}$ & $0.551_{-0.065}^{+0.056}$ & $0.211_{-0.037}^{+0.048}$\\
 $R^b$ [R$_\odot$] & $1.211_{-0.012}^{+0.012}$ & $1.217_{-0.014}^{+0.015}$ & $0.799_{-0.040}^{+0.034}$ & $0.532_{-0.041}^{+0.040}$\\
 $T^b_\mathrm{eff}$ [K]& $6552_{-44}^{+112}$ & $6281_{-39}^{+39}$ & $3580_{-71}^{+136}$ & $3077_{-79}^{+86}$\\
 $L^b_\mathrm{bol}$ [L$_\odot$] & $2.43_{-0.09}^{+0.12}$ & $2.07_{-0.06}^{+0.05}$ & $0.09_{-0.01}^{+0.02}$ & $0.023_{-0.007}^{+0.009}$\\
 $M^b_\mathrm{bol}$ & $3.81_{-0.05}^{+0.04}$ & $3.98_{-0.03}^{+0.03}$ & $7.33_{-0.26}^{+0.20}$ & $8.88_{-0.25}^{+0.26}$\\
 $M^b_V           $ & $3.83_{-0.05}^{+0.04}$ & $4.02_{-0.03}^{+0.04}$ & $8.89_{-0.46}^{+0.30}$ & $11.42_{-0.43}^{+0.45}$\\
 $\log g^b$ [dex]   & $4.33_{-0.01}^{+0.01}$ & $4.31_{-0.01}^{+0.01}$ & $4.37_{-0.01}^{+0.01}$ & $4.31_{-0.01}^{+0.01}$ \\
 \hline
 $\log$(age) [dex] &\multicolumn{4}{c}{$7.257_{-0.008}^{+0.016}$} \\
 $[M/H]$  [dex]    &\multicolumn{4}{c}{$-0.3019_{-0.051}^{+0.012}$} \\
 $E(B-V)$ [mag]    &\multicolumn{4}{c}{$0.0055_{-0.0030}^{+0.0074}$} \\
extra light $\ell_5$  [in {\em TESS}-band]&\multicolumn{4}{c}{$0.127_{-0.053}^{+0.093}$} \\
extra light $\ell_5$  [in SWASP-band]&\multicolumn{4}{c}{$0.034_{-0.038}^{+0.029}$} \\
$(M_V)_\mathrm{tot}^b$             &\multicolumn{4}{c}{$3.16_{-0.04}^{+0.03}$} \\
distance [pc]                &\multicolumn{4}{c}{$349.8_{-4.0}^{+6.3}$}  \\  
\hline
\end{tabular}

{\em Notes. }{$a$: Instantaneous, osculating orbital elements, calculated for epoch $t_0=2455413.533621$ (BJD); $b$: Interpolated from PARSEC isochrones; }
\end{table*}

\section{Physical parameters of the components}
\label{sec:par}

\begin{figure*}
\begin{center}
\includegraphics[width=0.49 \textwidth]{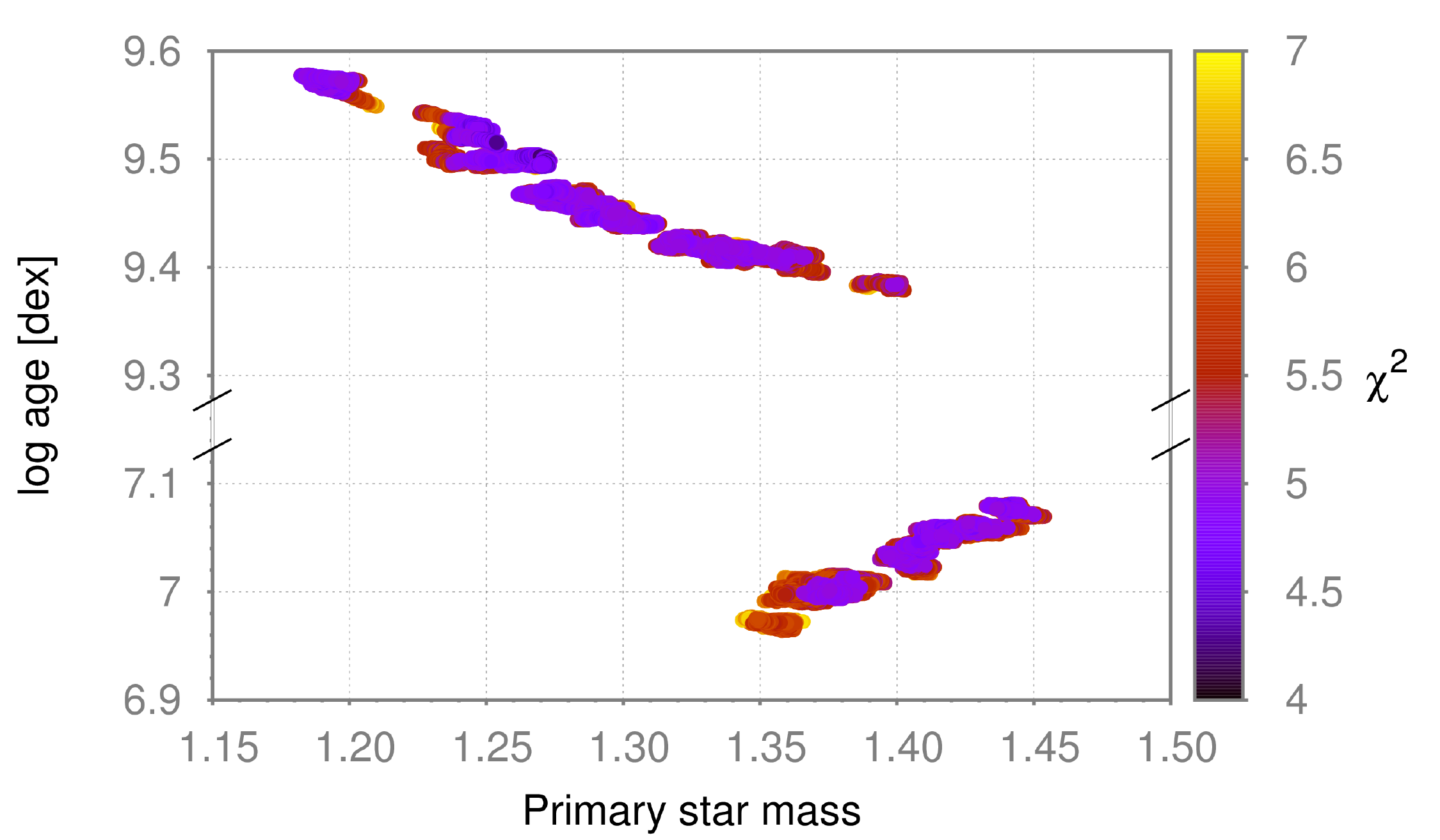}\includegraphics[width=0.49 \textwidth]{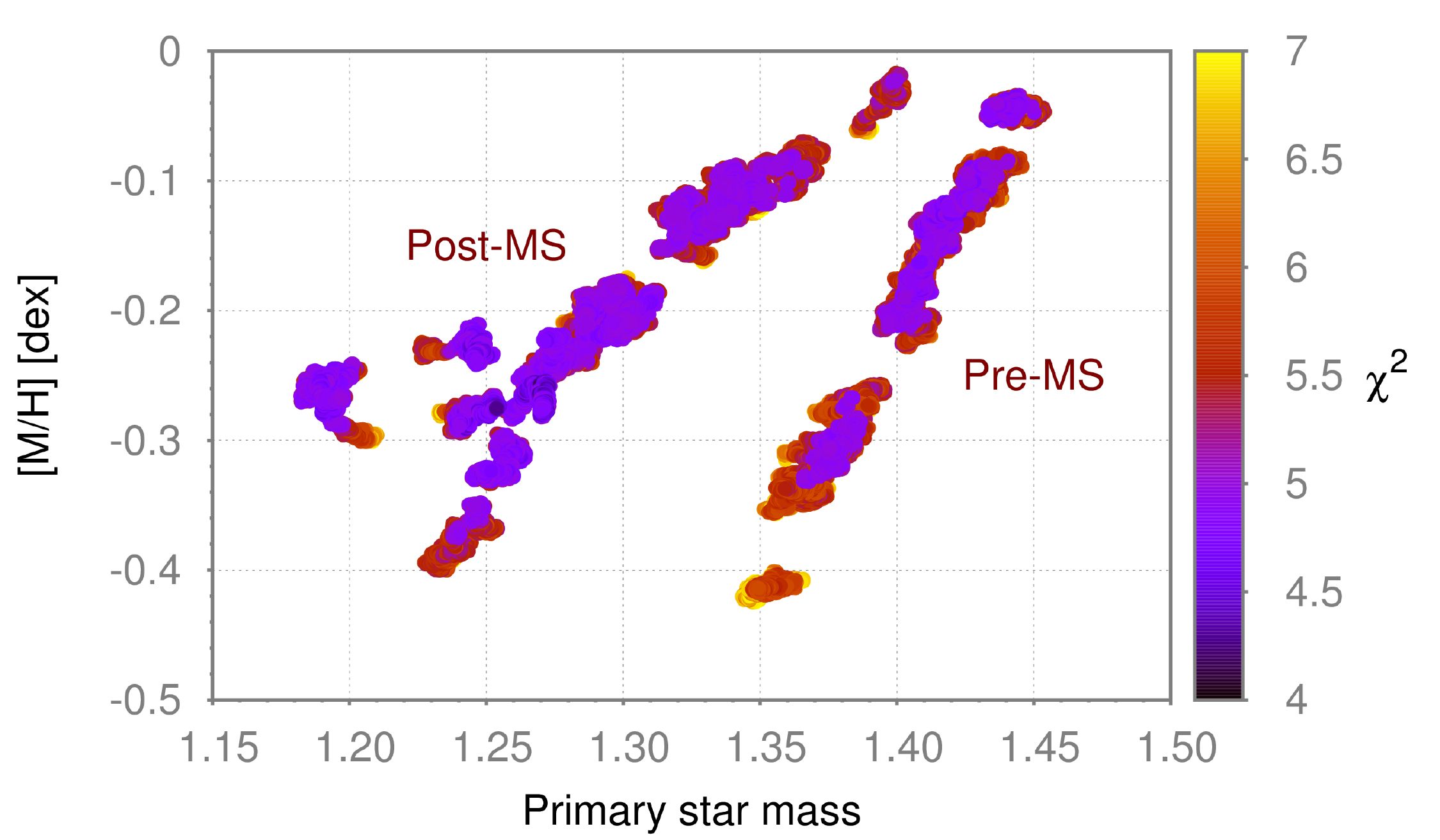}
\includegraphics[width=0.49 \textwidth]{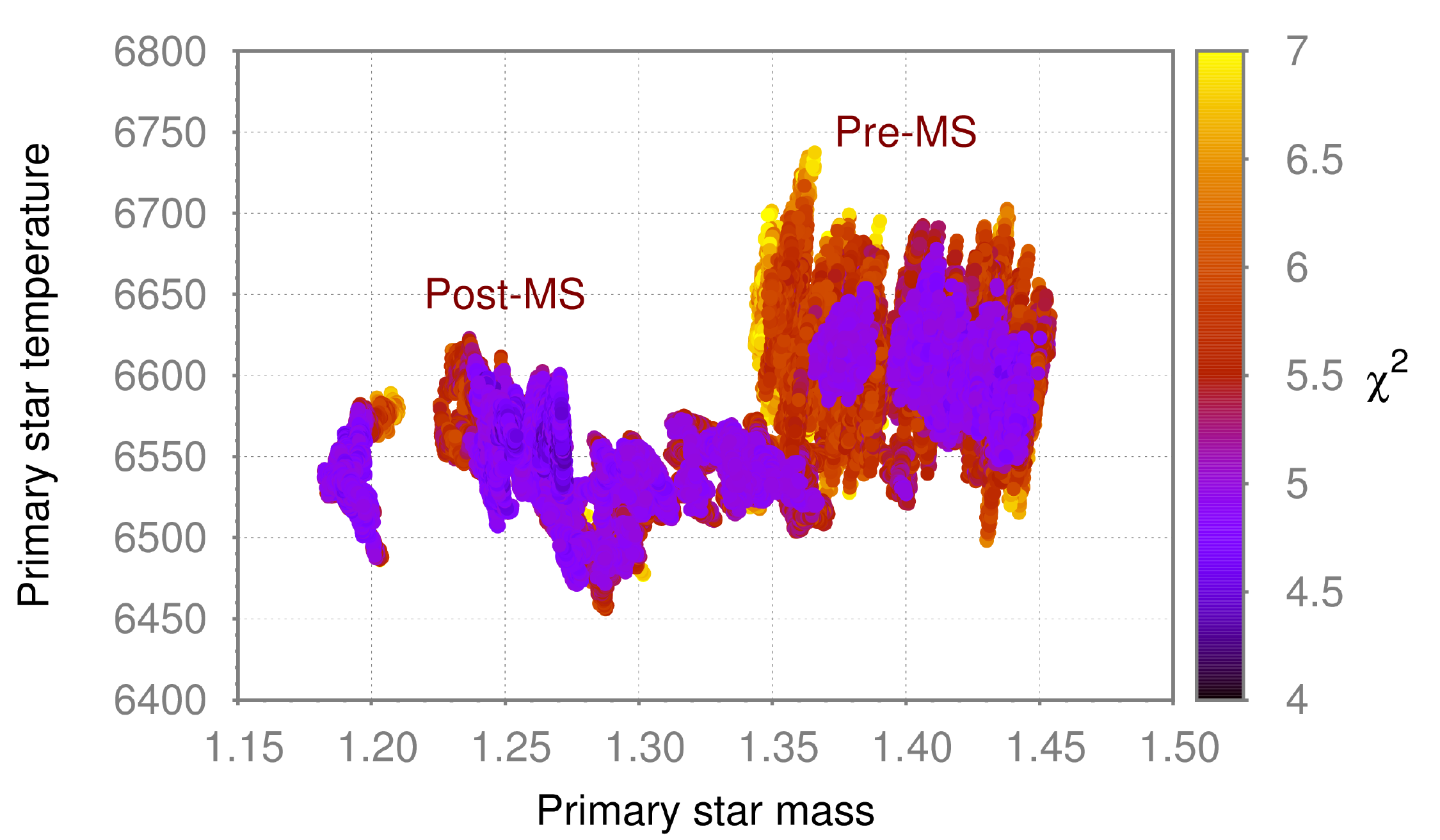}\includegraphics[width=0.49 \textwidth]{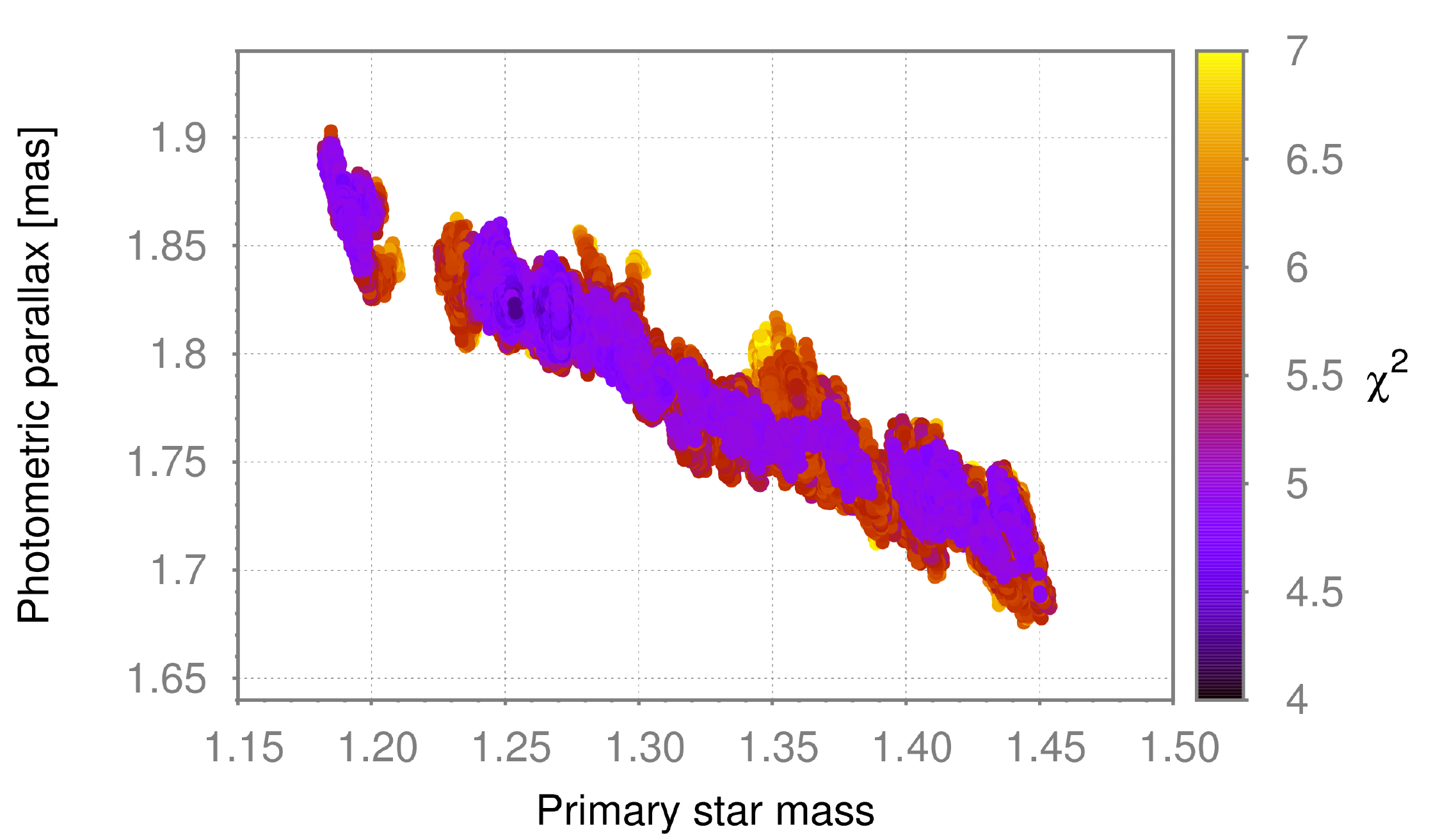}
\caption{Correlation plots for TIC\,167692429 among the primary star mass ($m_\mathrm{A}$) and its age ($\log\tau$), metallicity ($[M/H]$), effective temperature ($T_{\rm eff,A}$), and photometric parallax ($\varpi_\mathrm{phot}$).  Upper left and right panels are $\log\tau$ and $[M/H]$ vs.~$m_\mathrm{A}$, respectively. Lower left and right panels are $T_\mathrm{eff,A}$ and $\varpi_\mathrm{phot}$ vs.~$m_\mathrm{A}$, respectively.  The plotted points represent all the accepted MCMC trial steps both for the post-MS and pre-MS solutions. The color scale represent the $\chi^2$ value of each trial step and demonstrates that similarly low $\chi^2$ value-solutions can be obtained over a wide range of these parameters. Note the break in the y-axis in the upper left panel.} 
\label{fig:m1vslogagemh} 
\end{center}
\end{figure*}

\subsection{TIC\,167692429}

The preliminary stages of the analysis revealed that this system consists of non main-sequence components. The initial search amongst the PARSEC grids resulted in both pre- and post-MS isochrones. As was expected, we found that for lower metallicity values we obtained appropriate isochrones of lower mass stars (i.e., for a given initial mass, the more metal-rich a star is, the lower its effective temperature). Furthermore, we found, that for a given metallicity, the pre-MS isochrones, in general,  belonged to more massive stars than the corresponding post-MS ones. We used isochrone grids in the metallicity range $-1.182985\leq[M/H]\leq0.595166$. Within this range, our preliminary search resulted in probable primary masses within $0.80\,\mathrm{M}_\odot\lesssim m_\mathrm{A}\lesssim1.95\,\mathrm{M}_\odot$. As the effective temperatures of the EB members, and their relative radii, as well as the mass ratio are relatively well-known from the previous stages of the analysis, one can find that in this case, apart from additional light sources (i.e., the third component, being significantly less luminous), the total brightness of the system scales simply with the primary's mass according to $L_\mathrm{tot}\propto{m_\mathrm{A}}^{2/3}$. In such a way there is a direct relation between the primary's mass and the (photometric) distance to the system (see lower right panel of Fig.\,\ref{fig:m1vslogagemh}).

As was mentioned in the previous section, in this way we have found a significant discrepancy between the trigonometric distance derived from the Gaia DR2 parallax, and the photometric one obtained from our joint analysis. For the most massive solutions (i.e., for $m_\mathrm{A}\sim1.95\,\mathrm{M}_\odot$), our combined analysis---including the third star and the interstellar extinction---has resulted in a photometric parallax higher than the Gaia DR2 result at the $\approx3\sigma$ level.  Moreover, these most massive star scenarios belonged to the extreme metal rich stellar isochrones (i.\,e. $[M/H]\approx0.59$), while for TIC\,167692429, the TIC lists $[M/H]=-0.309829\pm0.0462466$. Regarding only the metal deficient (relative to the Sun) isochrones we find primary masses $m_\mathrm{A}\lesssim1.47\,\mathrm{M}_\odot$, which makes the distance discrepancy more significant.

This discrepancy perhaps can be resolved by the fact that the astrometric solutions used to produce the Gaia DR2 parallax do not take into account the wide outer binary nature of the systems.  Similar, or even much larger discrepancies, have been reported, e.g., by \citet{benedictetal18} who compared HST and Gaia Parallaxes and concluded that 8\% of their ``comparison sample of Gaia DR2 parallaxes have some issues with either target identification (high proper motion?) or binary motion.''  In our case the source of the discrepancy might be the $P_2\approx331$-day outer binary orbital motion. As one readily can find from Table\,\ref{tab: syntheticfit_TIC167}, $a_2\approx1.4$\,au. Taking into account that the outer mass ratio is $q_2\approx0.34$, the binary's center of mass orbit has a semi-major axis of $a_\mathrm{AB}\approx0.4$\,au.  According to our orbital solution the major axis of the outer orbit viewed nearly edge-on practically coincides with the node ($i_2\approx86\degr$; $\omega_2\approx0\degr$); therefore, the projected, near one-year-period orbital motion of the photocenter is practically a straight line segment having a length comparable ($\approx40$\%) to the trigonometric parallax, which might be co-measured with it.

\begin{figure*}
\begin{center}
\includegraphics[width=0.49 \textwidth]{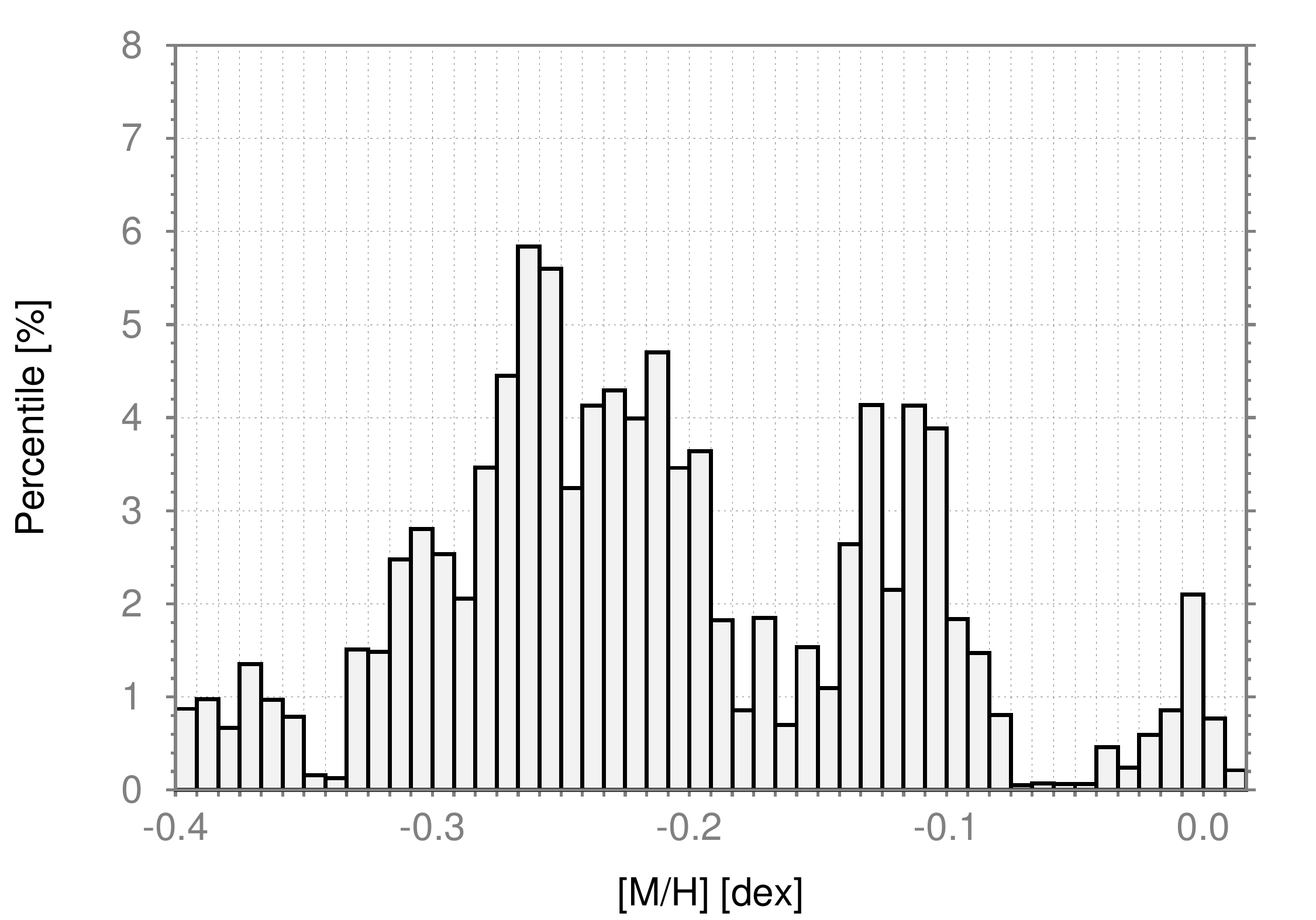}\includegraphics[width=0.49 \textwidth]{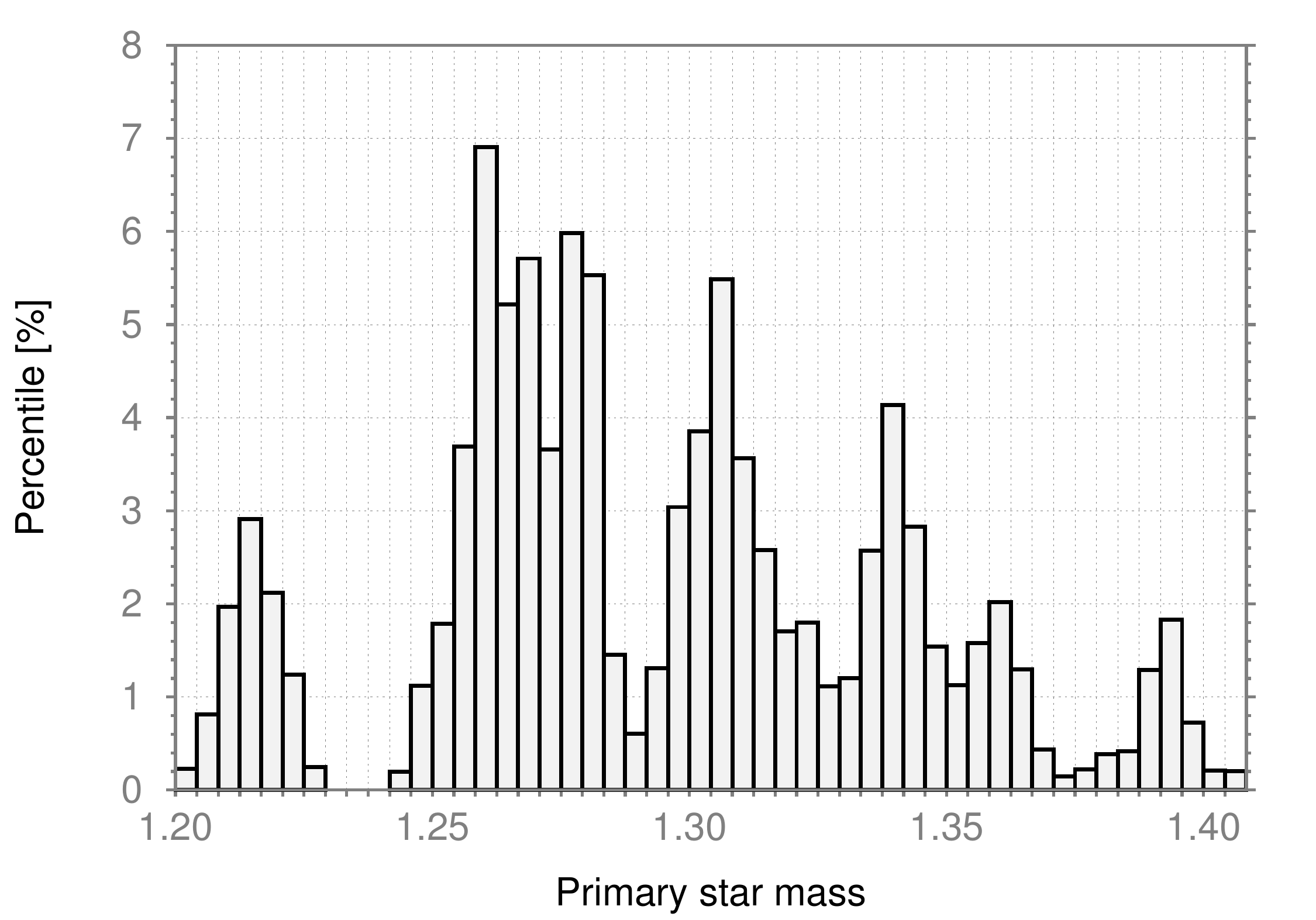}
\includegraphics[width=0.49 \textwidth]{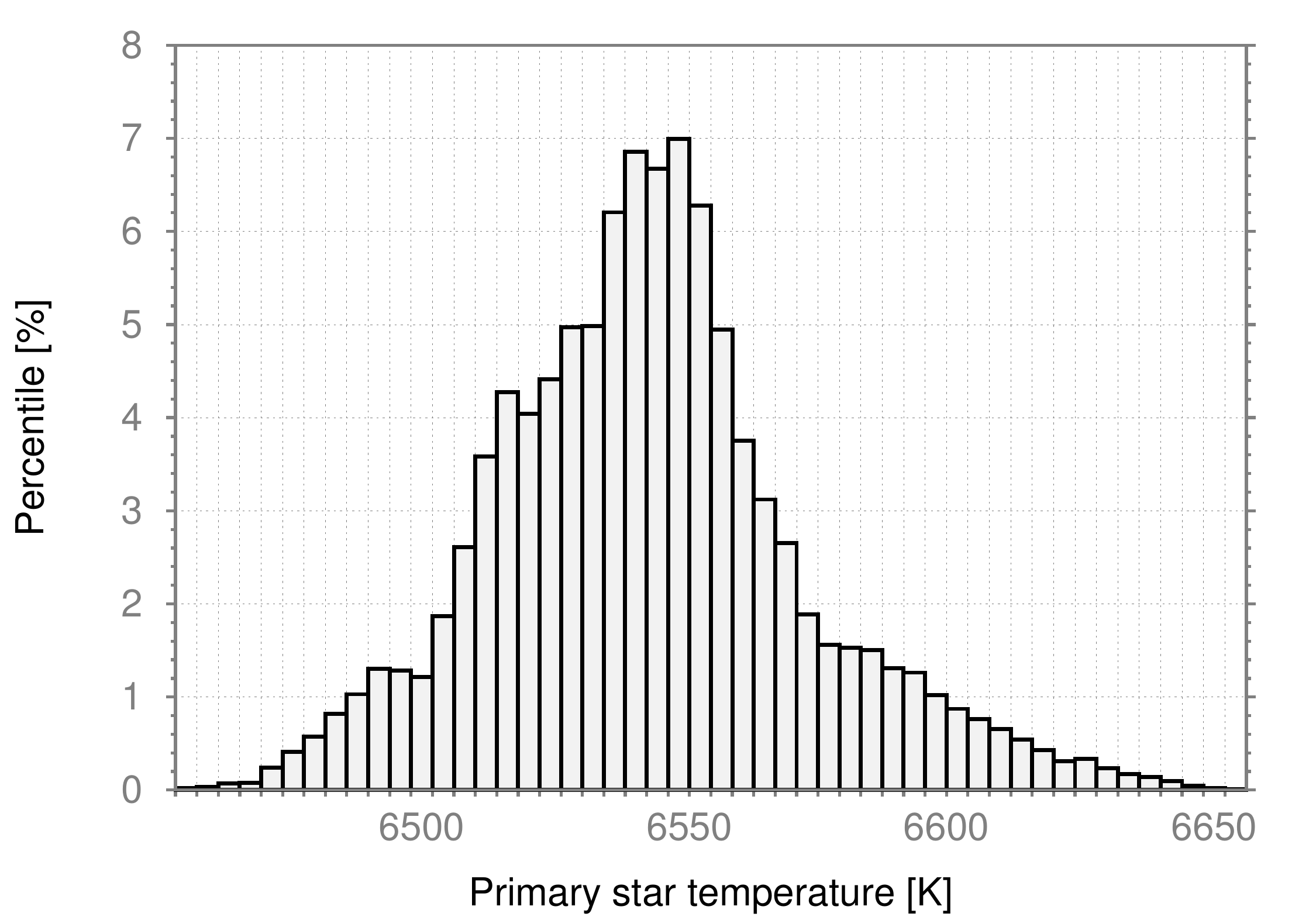}\includegraphics[width=0.50 \textwidth]{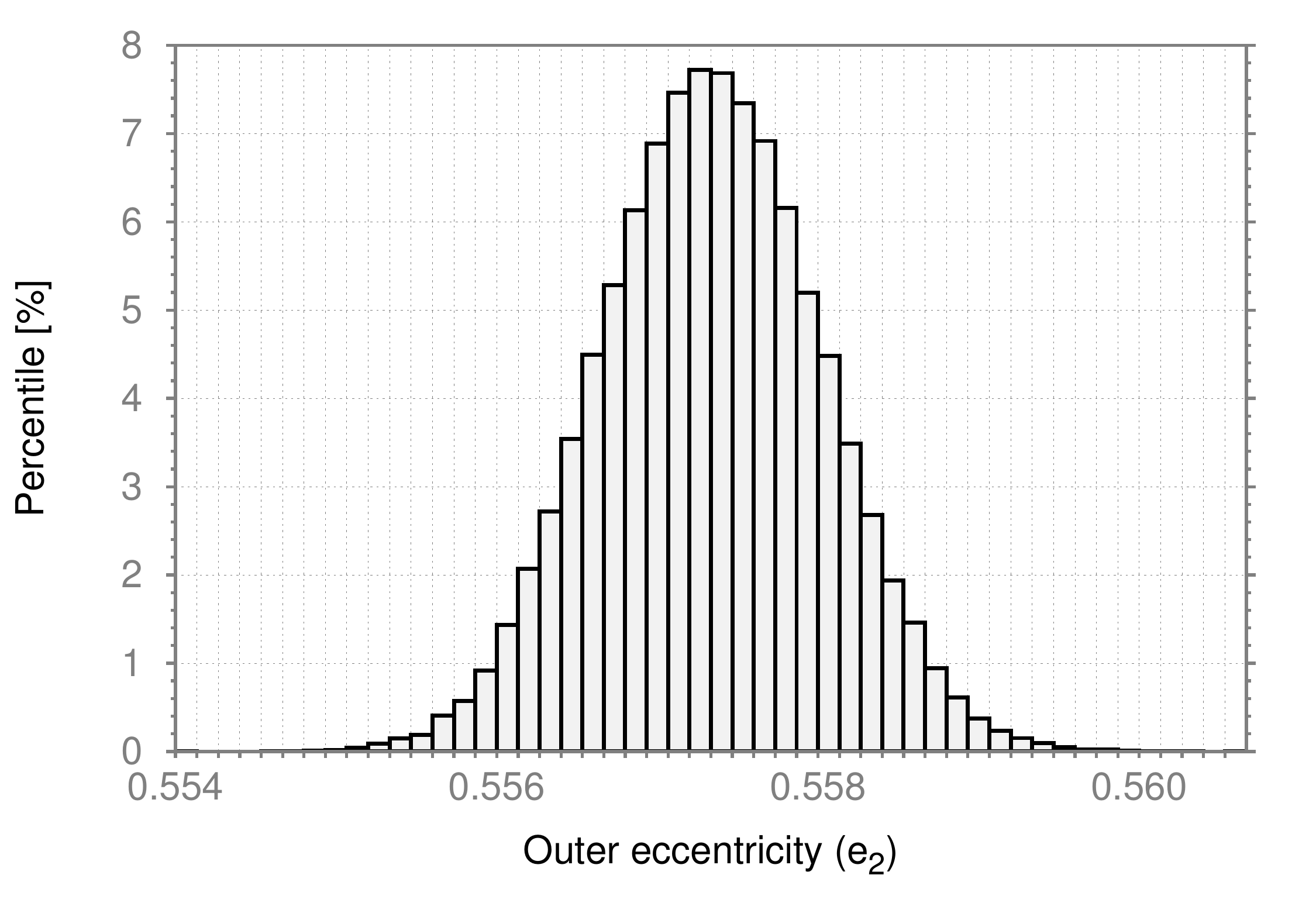}
\caption{Histogram plots illustrating the distributions of some parameters of the TIC\,167692429 system for the post-MS models. Upper row displays the highly non-gaussian distributions of the primary star's mass ($m_\mathrm{A}$) and  metallicity ($[M/H]$).  The bottom row shows the close to Gaussian distribution of the primary star's effective temperature ($T_{\rm eff,A}$), and the practically perfect Gaussian distribution of the outer eccentricity ($e_2$).  These distributions were chosen to be illustrative of similar distributions for the majority of the orbital elements.} 
\label{fig:histograms} 
\end{center}
\end{figure*}

In conclusion, unfortunately, in this case we cannot simply filter out the isochrones belonging to stellar masses that are inconsistent with the system's distance.  Instead, we selected isochrones with $[M/H]\leq0.0$, and looked for solutions within this metallicity constraint.  We initiated several MCMC chains both in the post- and the pre-MS domains.  In the upper panels of Fig.\,\ref{fig:m1vslogagemh} we show the $m_\mathrm{A}$ versus $\log\tau$ and $m_\mathrm{A}$ vs $[M/H]$ plots for both domains. These figures nicely illustrate the above mentioned strong interdependencies amongst stellar masses, ages and metallicities. Furthermore, one can see that the distributions of the appropriate stellar physical parameters are discontinuous, e.g., in the $m_\mathrm{A}$ vs $\log\tau$ and $m_\mathrm{A}$ vs $[M/H]$ planes. 

This fact is also well illustrated in Fig.\,\ref{fig:histograms} where one can readily see the distinctly structured nature of the distribution for $m_{\rm A}$. In particular, note the total lack of allowed solutions with primary mass $m_\mathrm{A}\sim1.24\,\mathrm{M}_\odot$. We claim that this is a real physical effect and not simply an artefact caused by insufficient MCMC sampling. We justify this claim by noting that even in step (iii) of our complex process, i.e., while searching for PARSEC isochrone grid elements that a priori fulfill some preconditions characteristic of the given system (i.e. primary star temperature, mass and temperature ratios, etc, see Sect.\,\ref{sec:dyn}), we found only a small number of appropriate grid elements at these primary masses (relative to other masses). And later, when we initiated additional runs setting the input values of $m_\mathrm{A}$, $[M/H]$, $\log\tau$ to lie directly in the gaps of Figs.\,\ref{fig:m1vslogagemh} and \ref{fig:histograms}, all these chains walked to the previously obtained islands of the a posteriori parameters. 

\begin{figure*}
\begin{center}
\includegraphics[width=0.49 \textwidth]{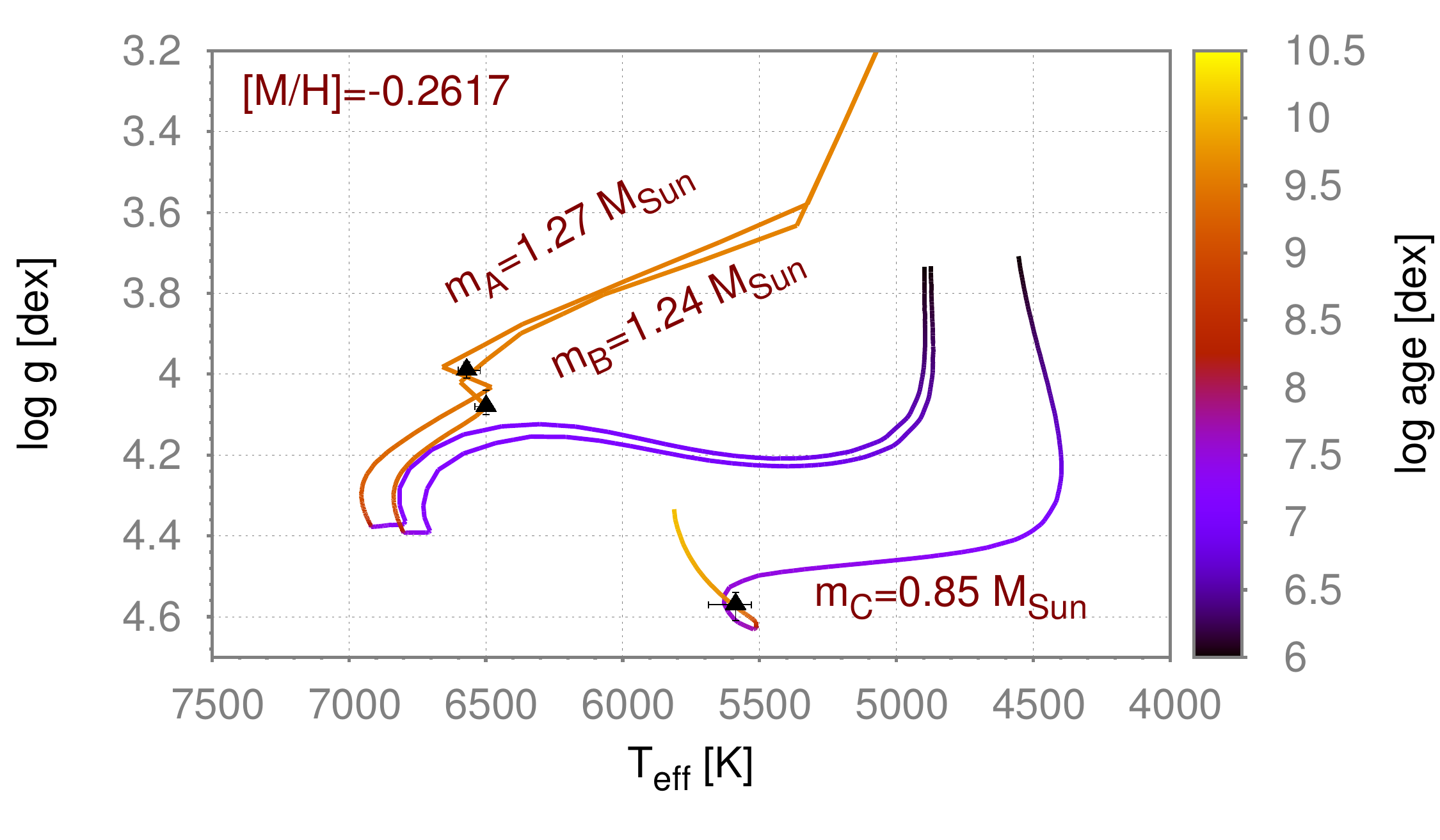}\includegraphics[width=0.49 \textwidth]{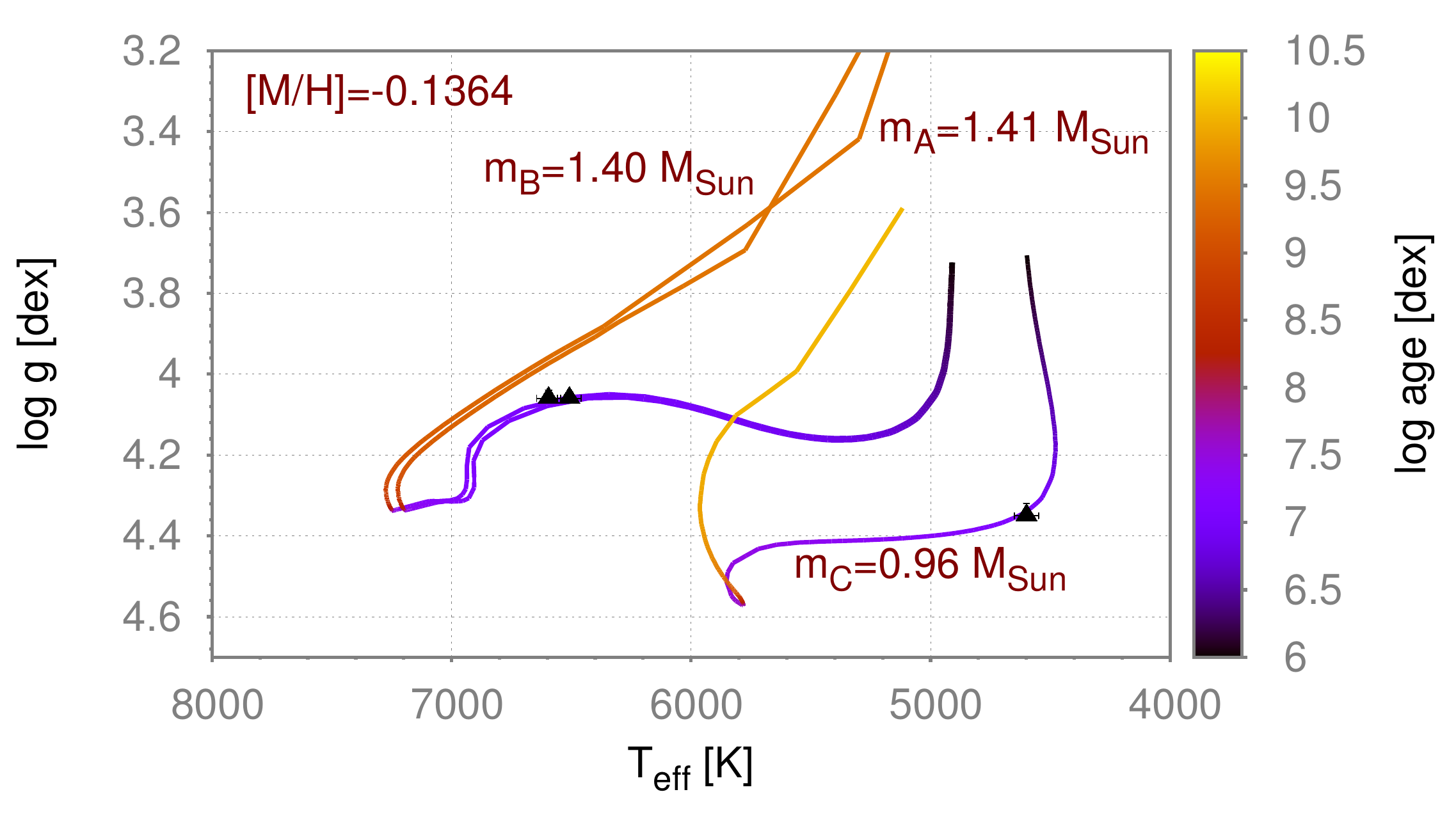}
\caption{$T_\mathrm{eff}$ vs $\log g$ PARSEC evolutionary tracks for the three components of TIC\,167692429 both for the evolved (left panel) and the pre-MS scenarios (right panel). The color scale denotes the age ($\log\tau$) of the stars at any given point along their evolution tracks. Black triangles mark the present locations of the three stars at ages $\log\tau\approx9.50$ and $\log\tau\approx7.05$ in the left and right panels, respectively.} 
\label{fig:isochrone1} 
\end{center}
\end{figure*}

The complex explanation for this fact is beyond the scope of this paper. Here we simply refer to the evolutionary tracks formed from the PARSEC isochrones (see Fig.\,\ref{fig:isochrone1}) where one can see that, e.g., in the case of the post-MS models for TIC\,167692429 all three stars are located at very rapidly and steeply varying parts of their evolutionary tracks.  We surmise that these `kinks' in the evolutionary tracks might cause there to be no combinations of coeval evolutionary tracks of the three stars (with given mass ratios) for specific primary masses which would produce the required combination of stellar parameters that match the observed data. 

In Table\,\ref{tab: syntheticfit_TIC167} we list our results  for both the post-MS and pre-MS domain. As one can see, apart from the ambiguity discussed above related to the triplets of (mass, age, metallicity), the two solutions are very similar.  In particular, the present orbital configuration and dynamics of our triple are very well determined by the observations and, in this sense, our results are conclusive. (The orbital and dynamical implications of the results will be discussed in the forthcoming section\,\ref{sec:orbprop}.)  From an astrophysical point of view, however, we cannot decide with certainty whether TIC\,167692429 is a young, pre-MS system or, conversely, it is an old, evolved system.  Though, the old, evolved scenario seems to be somewhat preferred in a statistical sense from our fits.  What is certain is that the inner binary is comprised of two F-type twin stars ($q_1=0.99\pm0.02$), and the distant, third star is a less massive, G-type object.  The locations of the three stellar components on their appropriate PARSEC evolution tracks for both solutions are plotted in Fig.\,\ref{fig:isochrone1}. It is interesting, at first glance, that the pre-MS solution results in hotter inner binary stars by $\Delta T_\mathrm{eff}\approx50-100$\,K relative to the post-MS solution (see, also, the lower left panel of Fig.\,\ref{fig:m1vslogagemh}). This difference corresponds to 2-3-$\sigma$ uncertainties of both solutions. We interpret this finding by noting the fact that, in the case of the pre-MS solution, the third stellar component is found to be cooler by $\approx1\,000$\,K, while its luminosity remains nearly the same as for the post-MS case. Therefore, the members of the inner binary must be hotter to counterbalance the flux excess of the cooler tertiary on the infrared tail of the cumulative SED curve. 

Despite the fact that our solutions have strongly degenerate dependencies on metallicities, masses, and ages, and the a posteriori distributions of some of the astrophyiscally most important parameters in the sample are far from Gaussian (see Fig.\,\ref{fig:histograms}), we estimated their uncertainties formally as if they had normal distributions. Specifically, we have simply chosen to report the median value of each parameter as well as integrating the usual percentiles in both directions from the median value. In this simplistic way we obtained, e.g., $3\%-5\%$ 1-$\sigma$ uncertainties in the stellar masses.  \footnote{Note, \citet{moedistefano15} have estimated similar uncertainties using a quite similar method to obtain fundamental parameters from combined lightcurve and isochrone analysis of EBs of the LMC.} The smaller uncertainties in the pre-MS case arise from the narrower mass-region which can produce acceptable solutions within the investigated domain of metal-deficient isochrones, i.e., $-1.18\leq[M/H]\leq0.0$ (again, see Fig.\,\ref{fig:m1vslogagemh}).

Our post- and pre-MS solutions give extinction-corrected photometric parallaxes of $\varpi_\mathrm{phot}=1.81_{-0.05}^{+0.02}$\,mas and $\varpi_\mathrm{phot}=1.74\pm0.02$\,mas, respectively, which are both substantially larger than Gaia's $\varpi_\mathrm{DR2}=1.41\pm0.03$\,mas. One can expect a resolution of this discrepancy when the DR3 edition of the binary-motion-corrected Gaia results is released in 2021.

Returning to the mass--age relations (Fig.\,\ref{fig:m1vslogagemh}), one can see that primary star masses in the range of $1.35\,\mathrm{M}_\odot\lesssim m_\mathrm{A}\lesssim1.41\,\mathrm{M}_\odot$ might pertain to both pre- and post-MS solutions.  As a consequence, if some future RV observations yield dynamical masses outside the above mass range, one will be able to decide immediately, whether the post- or the pre-MS scenario is valid. A similar statement can be made in the context of the expected very accurate Gaia DR3 distances. A trigonometric parallax of $\varpi_\mathrm{DR3}\gtrsim1.80$\,mas would clearly imply an evolved scenario, while $\varpi_\mathrm{DR3}\lesssim1.70$\,mas is expected to be found only for the case of the pre-MS solutions.

\subsection{TIC\,220397947}

\begin{figure}
\begin{center}
\includegraphics[width=0.49 \textwidth]{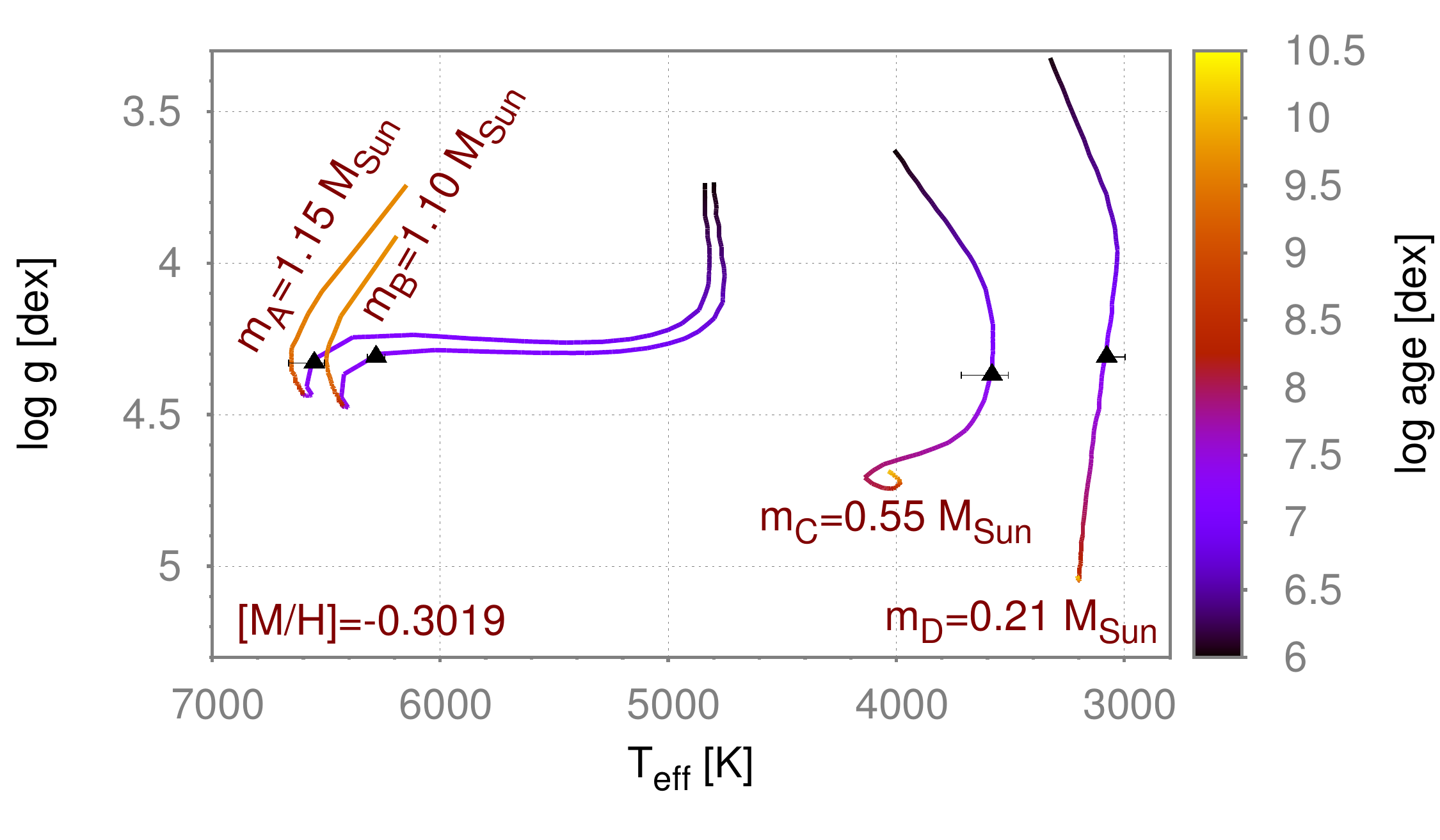}
\caption{$T_\mathrm{eff}$ vs $\log g$ PARSEC evolutionary tracks for the four components of TIC\,220397947 according to our pre-MS model. The color scale denotes the age ($\log\tau$) of the stars at any point along their evolution tracks. Black triangles mark the present locations of the four stars at age $\log\tau\approx7.26$.} 
\label{fig:isochrone2} 
\end{center}
\end{figure}

As discussed above, for this system we took into account the Gaia DR2 parallax, applying a Gaussian prior to the photometrically obtained parallax. We found that in the neighbourhood of the catalog's metallicity values of $[M/H]=-0.668558\pm0.0612805$ our solutions led to photometric parallaxes that were too large compared to Gaia's result and, therefore, these strongly metal deficient solutions were highly penalized by the Gaussian prior. We found that moderately metal-deficient isochrones offer solutions which are in accord with the Gaia distance.  A bit surprisingly, however, our MCMC parameter searches favoured very young stellar ages, i.e., pre-MS star solutions instead of evolved star scenarios (see Table\,\ref{tab: syntheticfit_TIC220} and Fig.\,\ref{fig:isochrone2}, as well). In our understanding, this arises from that fact that the inner mass ratio ($q_1$), which is determined chiefly by the ETV curve, and the temperature ratio ($T_\mathrm{eff,B}/T_\mathrm{eff,A}$), which primarily sets the primary-to-secondary eclipse depth ratio, were slightly discrepant for the evolved star solutions. In other words, in the case of post-MS solutions, when the mass ratio ($q_1$) was found from the ETV fit, the secondary eclipses were too deep relative to the primary ones (i.e., $\chi^2_\mathrm{LC}$ became larger), while for the correct temperature ratio and, therefore proper eclipse depths, the ETV residuals were too large (i.e., $\chi^2_\mathrm{ETV}$ penalized the solution). 

Note, the circular inner orbit does not contradict the inferred very young age of the system.  As was shown by \citet{zahnbouchet89}, orbits of late type stars with $P\lesssim7-8$\,d are expected to circularize by the end of their first million years of pre-MS evolution.

\section{Orbital properties and dynamical evolution}
\label{sec:orbprop}

In contrast to the stellar ages and masses, the dynamical properties of both systems are robustly determined.

\subsection{TIC 167692429}

\begin{figure*}
\begin{center}
\includegraphics[width=0.49 \textwidth]{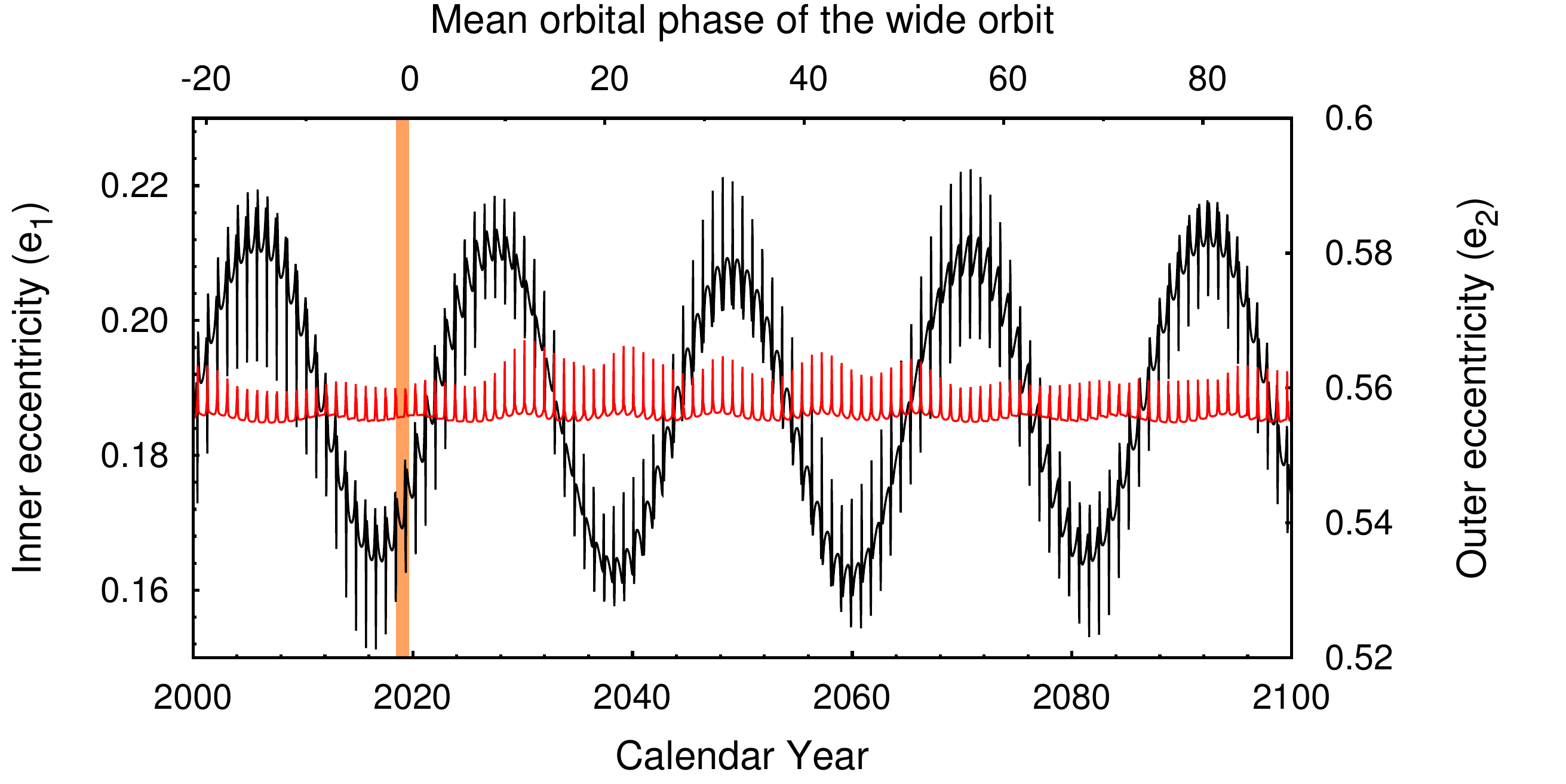}\includegraphics[width=0.49 \textwidth]{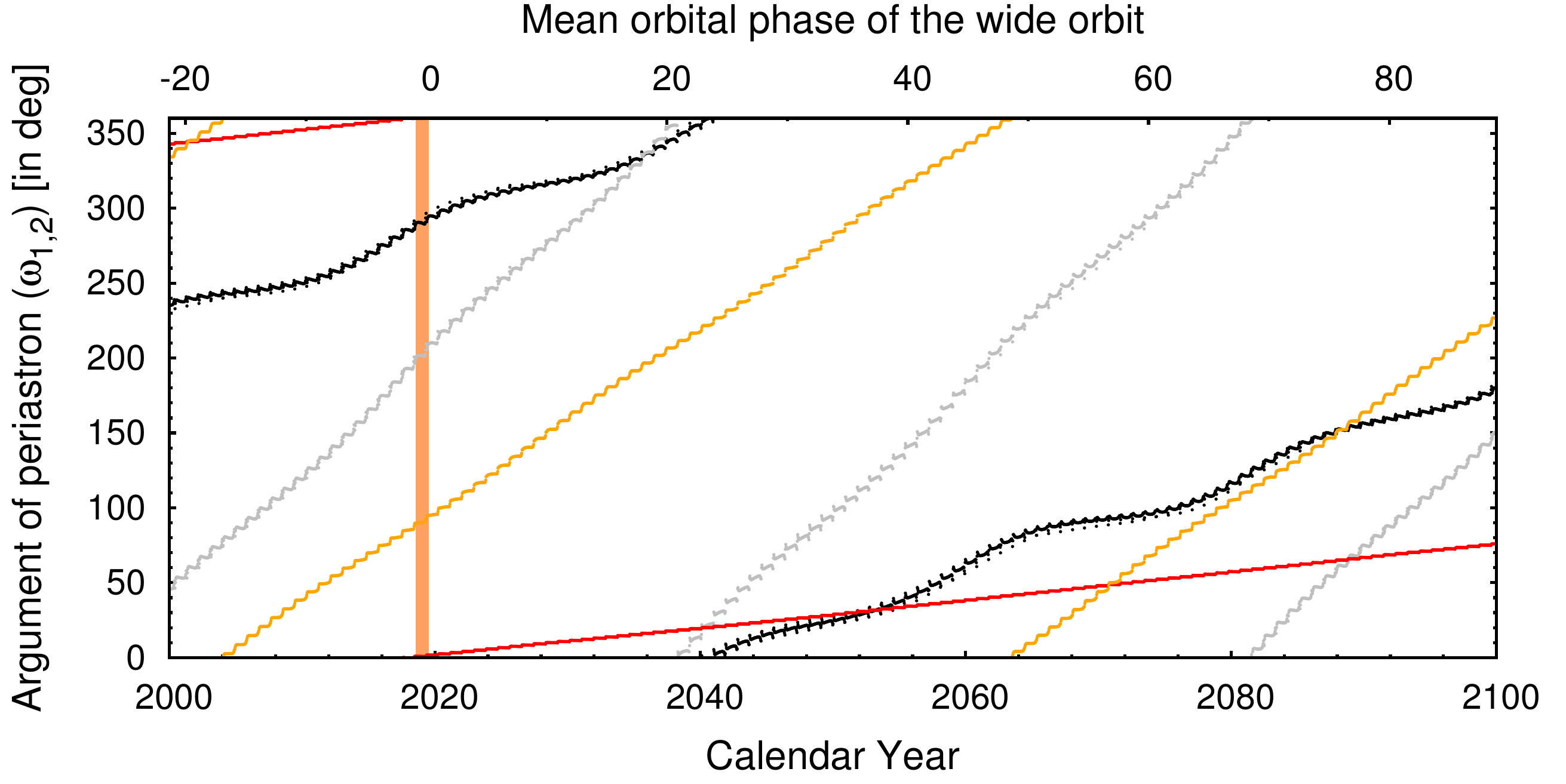}
\includegraphics[width=0.49 \textwidth]{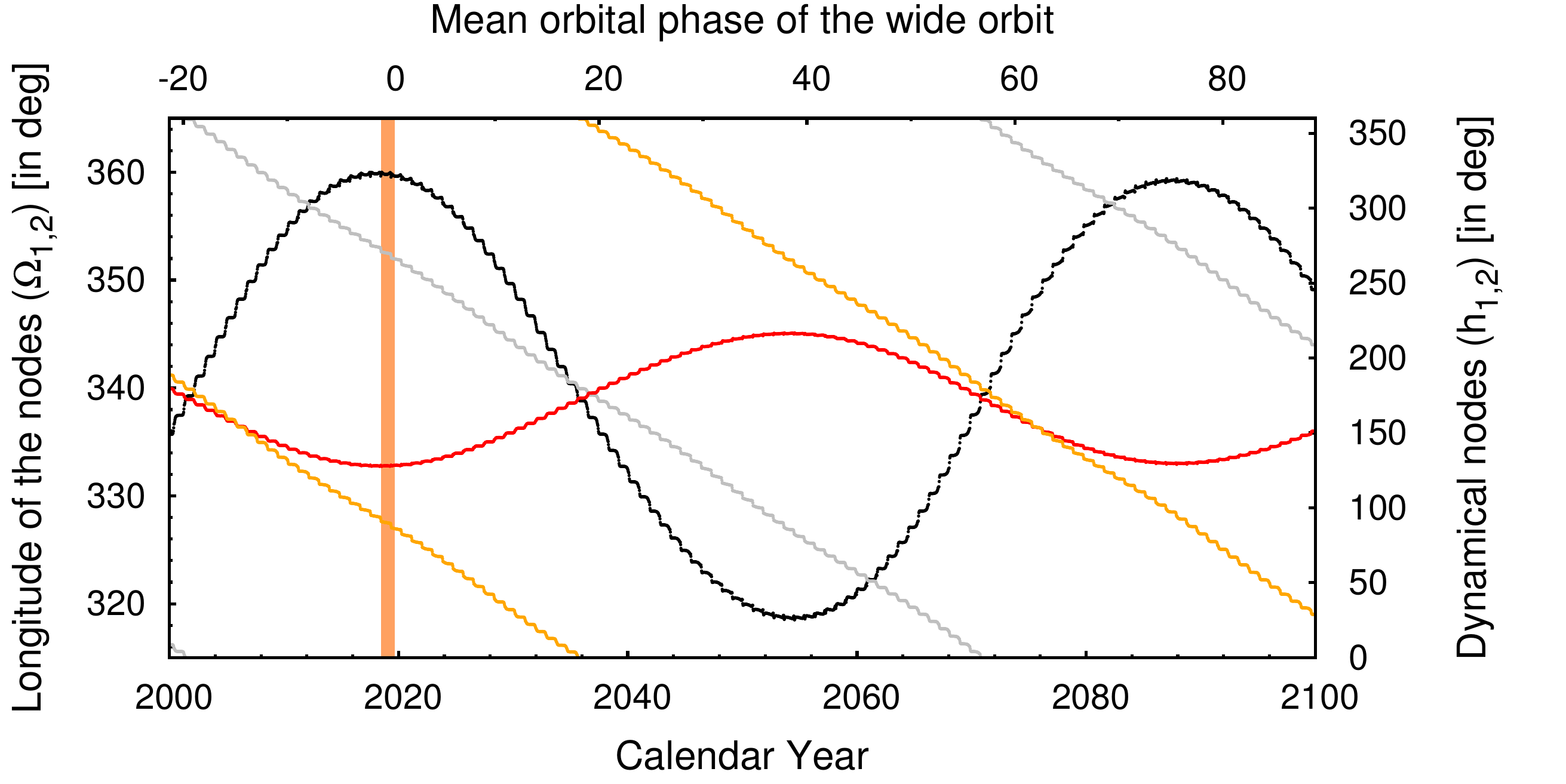}\includegraphics[width=0.49 \textwidth]{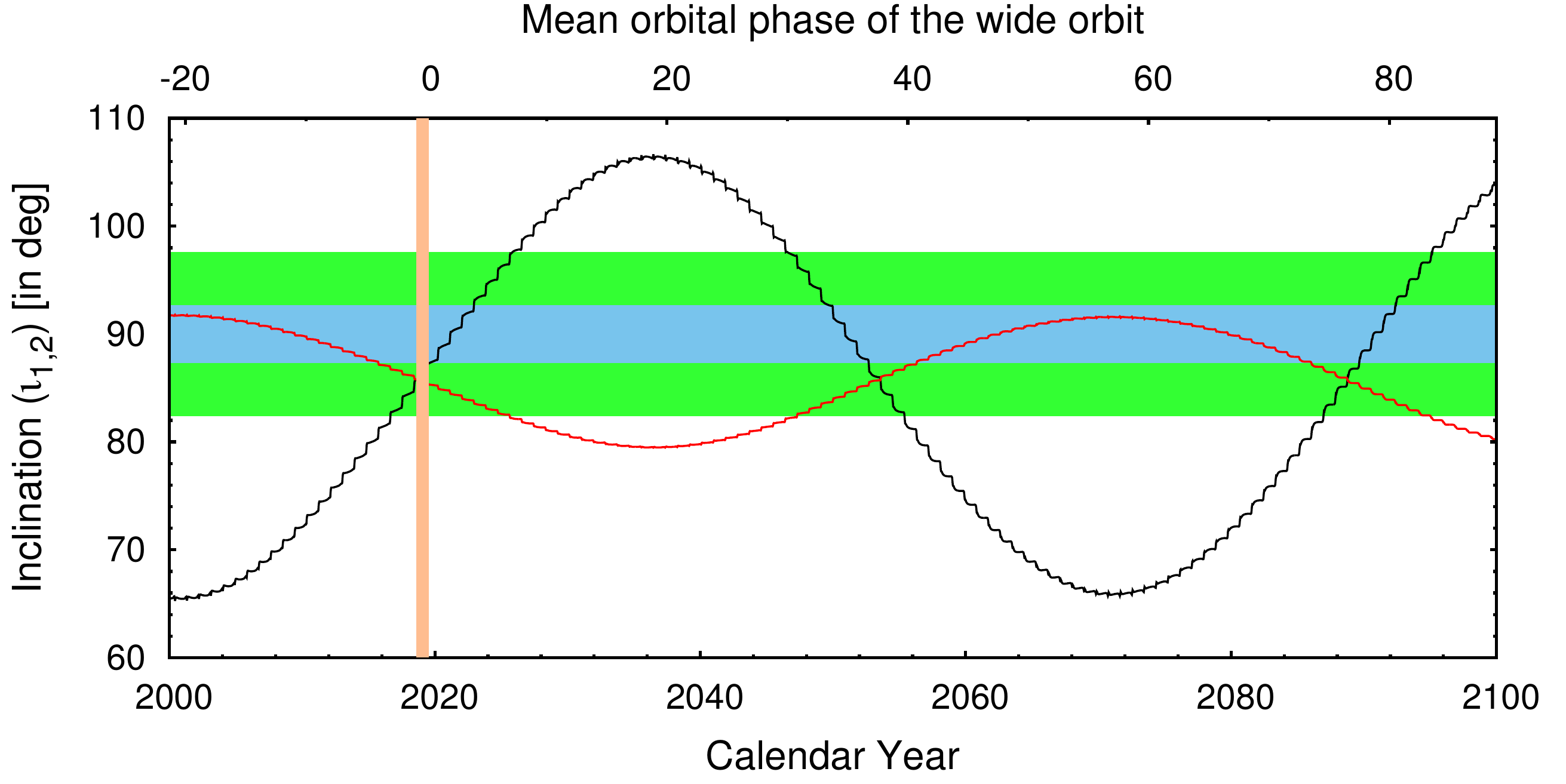}
\caption{Variations of different instantaneous (osculating) orbital elements for TIC\,167692429, obtained via numerical integrations. (The orbital elements were sampled nearly at the same orbital phases of the inner binary during each of its cycles.) The vertical orange shaded region in each panel marks the interval of the {\em TESS} observations. {\it Upper left, eccentricities:} Inner and outer orbits (black and red, respectively).  {\it Upper right, arguments of periastron:} Observable (measured from the intersection of the tangential plane of the sky and the respective orbital plane) and dynamical (measured from the intersection of the two orbital planes) for the inner (black -- observable; grey -- dynamical) and outer orbits (red -- obs.; orange -- dyn., respectively). {\it Bottom left,  longitudes of the nodes:} Observable (measured from an arbitrary starting point toward the intersection of the tangential plane of the sky and the respective orbital plane along the tangential plane of the sky) and dynamical (measured from the intersection of the invariable plane with the tangential plane of the sky toward the intersection of the two orbital planes along the invariable plane) for the inner (black -- observable; grey -- dynamical) and outer orbits (red -- obs.; orange -- dyn., respectively).  {\it Bottom right, inclinations:} Inner and outer binaries (black and red curves, respectively). The wide green-shaded horizontal area denotes the inclination ($i_1$) domain of the inner binary where regular eclipses can occur. The narrower blue-shaded area stands for the outer inclination ($i_2$) domain for possible outer eclipses. (See text for details.)}
\label{fig:orbelements_numint} 
\end{center}
\end{figure*}

This triple has a mutual inclination of $i_\mathrm{m}\approx27\degr$, which remains below the high eccentricity excitation Lidov-Kozai regime \citep[i.e., $141^\circ \gtrsim i_{\rm m} \gtrsim39\degr$, see e.\,g.,][]{lidov62,kozai62,naoz16} and excites only small, but rapid eccentricity oscillations with a full-amplitude of $\Delta e_1\approx0.05$ and a period of $P_\mathrm{e_1}\approx8000$\,d.  Therefore, one can conclude that the present configuration of the system is stable. We also confirmed this conclusion with a 10 million-year-long numerical integration, which did not show any dramatic variations in the orbital parameters. Therefore, we restrict our discussion only to some short-term, (partly) observational related facts. We plot the variations of some of the orbital elements during the first century of the recent milllenium in the four panels of Fig.\,\ref{fig:orbelements_numint}. Besides the above mentioned cyclic, apse-node timescale eccentricity variations, the spikes around the periastron passage of the outer orbit are also clearly visible (upper left panel). One can expect to detect these $\sim20$-yr period eccentricity cycles via radial velocity follow up observations. Furthermore, the dominant dynamically forced apsidal motion is also clearly visible (upper right panel). We plot the variations of both the observable arguments of periastron, i.e., the angle between the intersection of the respective orbital plane with the tangential plane of the sky, and the periastron point of the given orbit, and its dynamical counterpart, i.e., a similar angle measured between the intersection of the two orbital planes and periastron. While the former angles can be directly obtained from both radial velocity measurements and ETV and lightcurve analyses, the latter ones have an important role in the dynamical evolution of the system \citep[see, e.\,g.][for a more detailed discussion about the different effects of the observable and dynamical arguments of periastrons and nodes]{borkovitsetal07,Borkovits2015}. 


\begin{figure}
\begin{center}
\includegraphics[width=0.49 \textwidth]{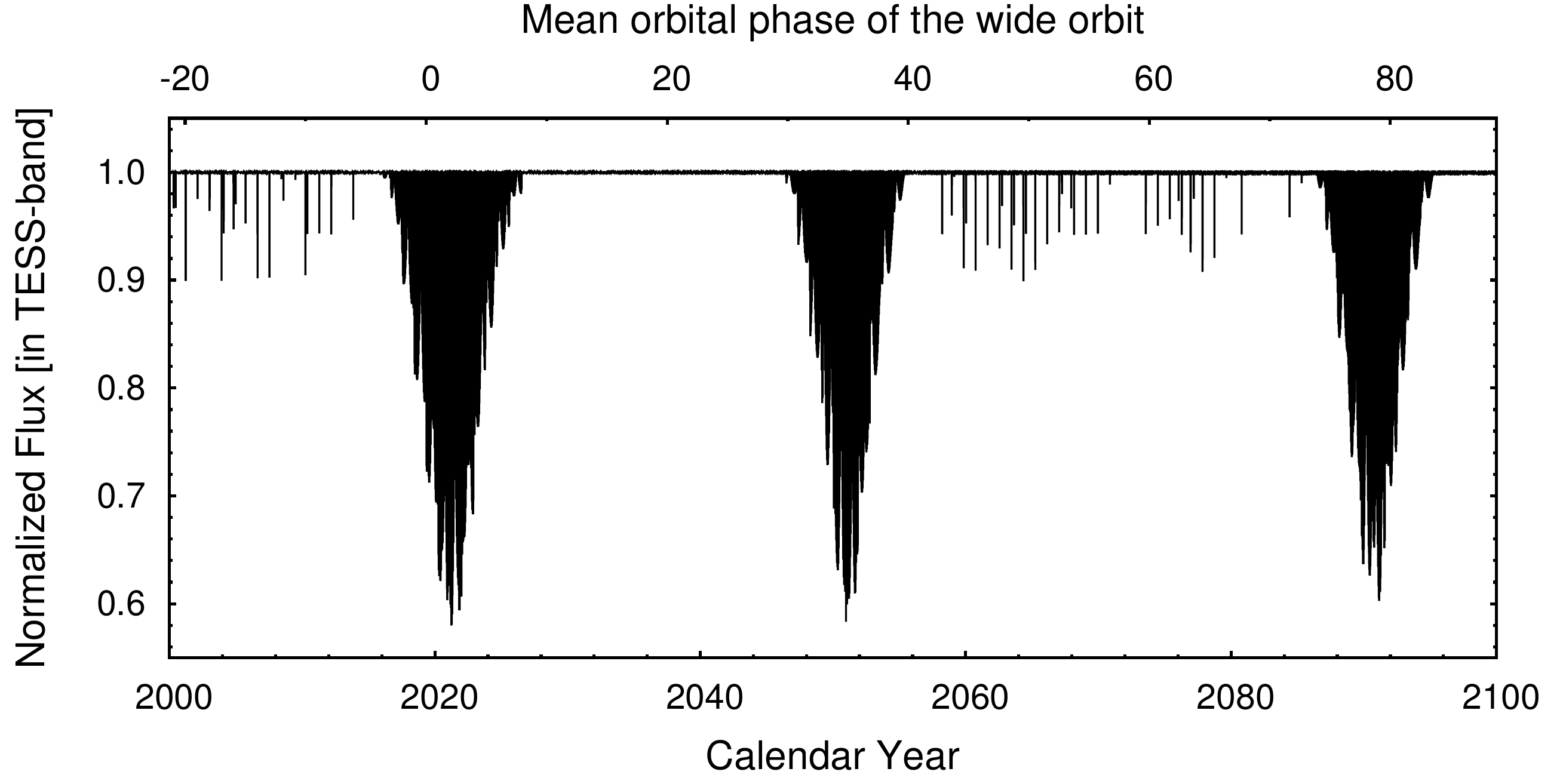}
\caption{Photodynamical model lightcurve of TIC\,167692429 for the first century of the present millennium. During one cycle of an $\sim70$\,yr-long precession cycle there are two $\sim11$\,yr-long intervals when the inner binary exhibits eclipses with continuously varying eclipse depths. Furthermore, interestingly, during the longer gaps between the regular binary eclipses, the system is subjected to outer eclipses, i.e., events when the outer, third star eclipses one or both members of the inner binary or, is eclipsed by them.}
\label{fig:T167692429lc100} 
\end{center}
\end{figure}

As a consequence of their non-coplanarity, both orbital planes precess with a period of about $P_\mathrm{node}\approx70$\,yrs. During this interval, the dynamical nodes regress by 360\degr on the invariable plane (see bottom left panel of Fig.\,\ref{fig:orbelements_numint}), while the normals to the inner and outer planes move along cones with half angles equal to the dynamical inclinations ($i_\mathrm{dyn1,2}\approx21\degr$ and $\approx6\degr$, respectively) of the two orbits. The consequent variations of the observable inclinations are plotted in the bottom right panel, while the most spectacular observational effect of this precession is shown in Fig.\,\ref{fig:T167692429lc100}. Perhaps the most interesting feature of this triple is that, for the almost edge-on invariable plane ($i_\mathrm{inv}\approx85\fdg6$), the system is also subject to outer eclipses during certain intervals. Interestingly, by chance, the intervals of the regular inner, and the more or less random outer, eclipses do not overlap each other. The last period of outer eclipses has ended in 2014 just one year before the start of the recent inner eclipsing episode at the end of 2015. Similarly, the forthcoming outer eclipse is expected in 2056, while the next cycle of regular, inner eclipses is predicted to finish in 2055. 

\begin{figure*}
\begin{center}
\includegraphics[width=0.49 \textwidth]{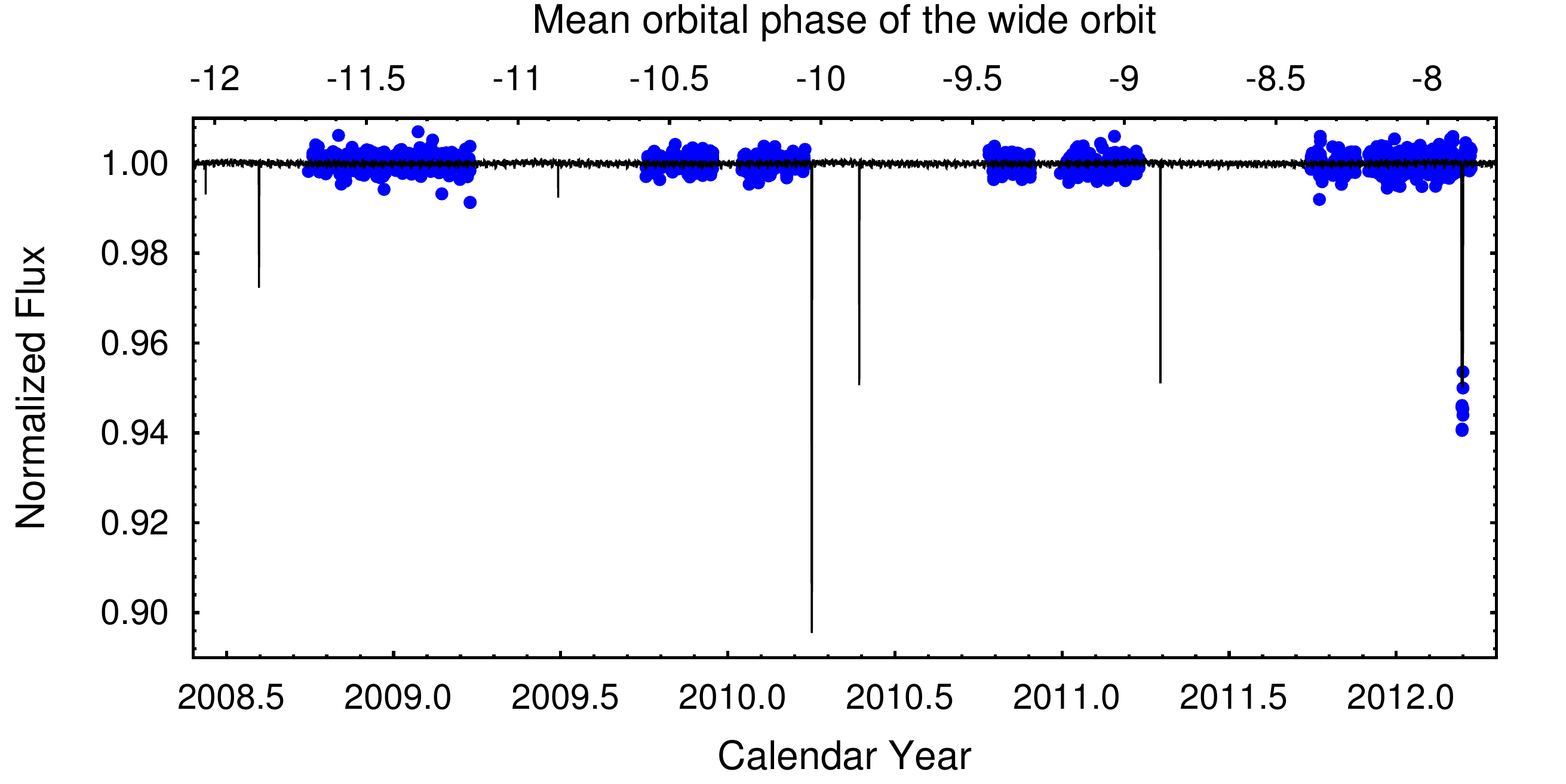}\includegraphics[width=0.49 \textwidth]{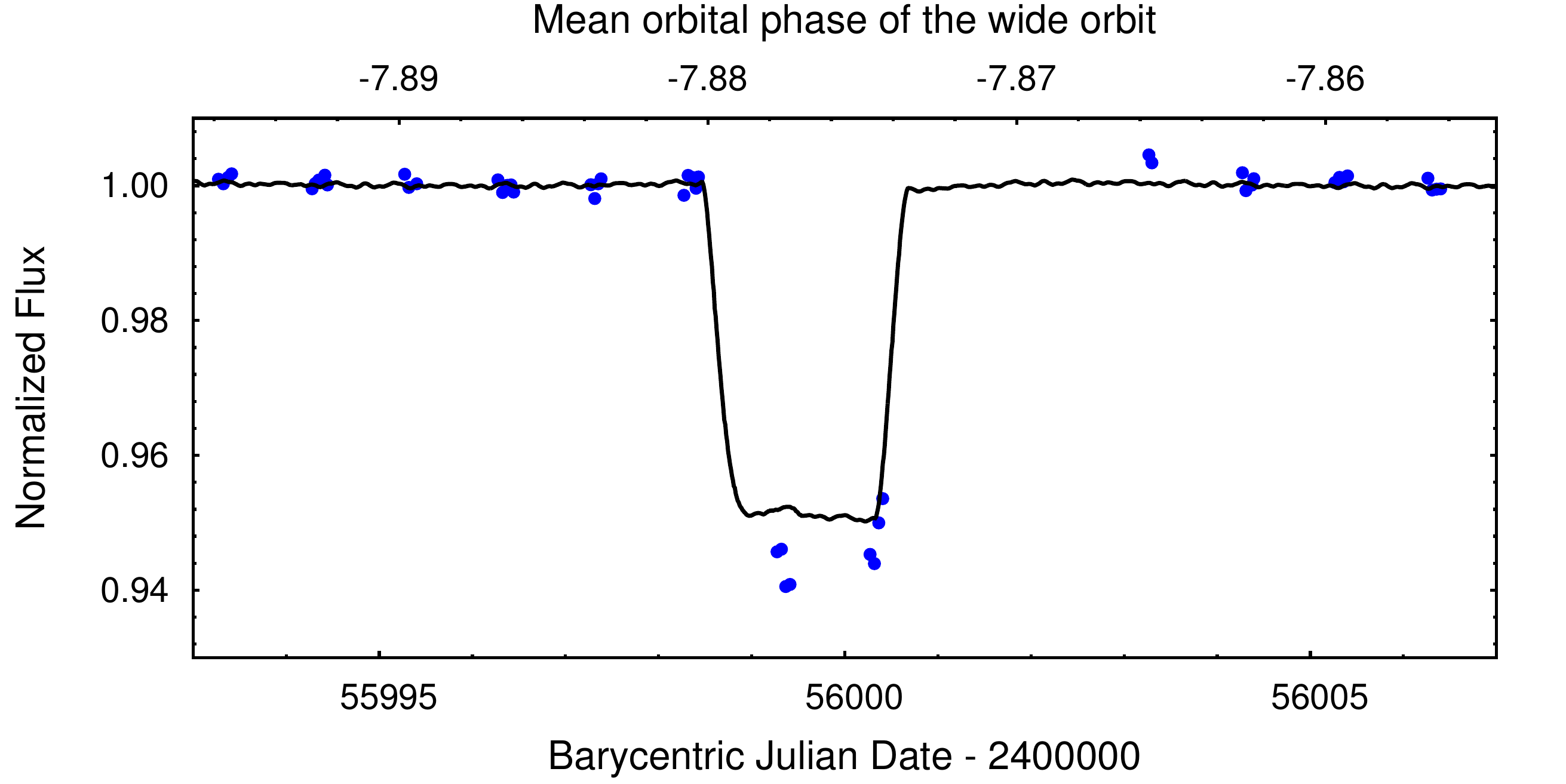}
\caption{WASP observations of TIC\,167692429 in one-hour bins (blue circles) together with the photodynamical model lightcurve (black line). It can clearly be seen in the {\em left panel} that all but one of the outer eclipses are located within seasonal gaps of the WASP measurements. A zoom-in of the only extra dimming observed during the WASP measurements is shown in the {\em right panel}. The corresponding section of the photodynamical model lightcurve confirms that this dip is probably a chance observation of an extra eclipse in this triple.}
\label{fig:T167692429lcSWASP} 
\end{center}
\end{figure*}

Realizing that, according to our photodynamical solution, the target might have produced outer eclipses during the interval of the WASP observations we checked the photodynamical model lightcurve against the WASP observations. In Fig.\,\ref{fig:T167692429lcSWASP} we plot the (1-hour binned) WASP observations together with the photodynamical model. We found that all but the last outer eclipse during that interval fell into seasonal gaps (see left panel of Fig.\,\ref{fig:T167692429lcSWASP}). The last event, however was found to be very close to the only extra dimming observed by the WASP cameras on the nights of 12 and 13 March, 2012, which previously was thought to be an artefact.  Therefore, as a very last step, we added the WASP lightcurve to the complex photodynamical, SED, and isochrone parameter search process, to refine our solution, and we were actually able to find sets of the initial parameters which led to solutions in accord with the location of the dimming observed by WASP (right panel of Fig.\,\ref{fig:T167692429lcSWASP}). 

Note, however, that our extended MCMC runs in this last stage were unable to reproduce the depth of this extra dimming event perfectly. All the accepted model parameter sets resulted in a systematically shallower extra eclipse with a discrepancy of $\sim0.01-0.015$\,mag. This might result from slightly discrepant (i) model ratios of the stellar surface brightnesses; (ii) other parameters such as the size of star C (which was fully eclipsed by the inner binary during that event) relative to the A-B binary members; or (iii) some of the dynamical parameters which could result in somewhat inaccurately modelled orbital perturbations going back six years. One should keep in mind, however, that there are only 6 points out of a total of $\sim1700$ lightcurve points which were used in the lightcurve modeling part of our fitting process. And, only these six points carry any direct information, for example, about the surface brightness and radius ratio of star C relative to the inner A-B binary members. Consequently, we cannot expect a perfect fit from such minimal information content. Despite this minor discrepancy, however, we can conclude, that the WASP observations confirm the former extra-eclipsing nature of our target and, of course, this fact makes our solution more robust.

\subsection{TIC 220397947}

In contrast to the system above, this target has shown constant eclipse depths not only during the 10-month interval of {\em TESS} measurements, but similarly deep regular eclipses were also observed continuously during the four seasons of the WASP observations. This fact suggests that the inner triple system should be very flat. Not surprisingly, our photodynamical model has resulted in a mutual inclination of $(i_\mathrm{mut})_\mathrm{AB-C}\approx0\fdg6$. On the other hand, however, as mentioned earlier, we found a clear discrepancy between the occurrence times of the WASP and {\em TESS} eclipses. In order to resolve this discrepancy, we assumed that TIC\,220397947 is indeed an--at least--quadruple system with a hierarchical of 2+1+1 structure. The outmost orbital solution is, however, quite ambiguous and, even the presence of the fourth component remains questionable. Fortunately, apart from the time shift in the moments of the eclipses, the presence or absence of this low-mass stellar component affects only weakly both the present-day astrophysical and dynamical parameters of this system.

The situation is reminiscent of the case of the recently discovered quadruple system EPIC\,212096658 \citep{Borkovits2019b}. That system consists of a similarly flat and compact ($P_2/P_1=59/2.9\approx20.3$ vs  $P_2/P_1=77/3.6\approx21.4$, for the {\em K2} and {\em TESS} systems, respectively) inner triple subsystem, where the innermost EB is also formed by two stellar twins ($q_1=0.98$ vs $0.95$), though the stars of the former EB themselves are significantly less massive.  Moreover, both systems consist of nearly circular inner orbits. The fourth, outermost, less massive component of EPIC\,212096658 was found through the systematic deviations in both the systemic radial velocity of the triple system and the ETV residuals. In that case, however, the RV observations have covered more than 4.5 outermost orbital cycles and, therefore, the presence of the fourth star seems to be certain. In the present situation future RV and/or eclipse timing observations are necessary to judge the four-body hypothesis, and refine the orbital parameters of the widest orbit.  

\begin{figure*}
\begin{center}
\includegraphics[width=0.49 \textwidth]{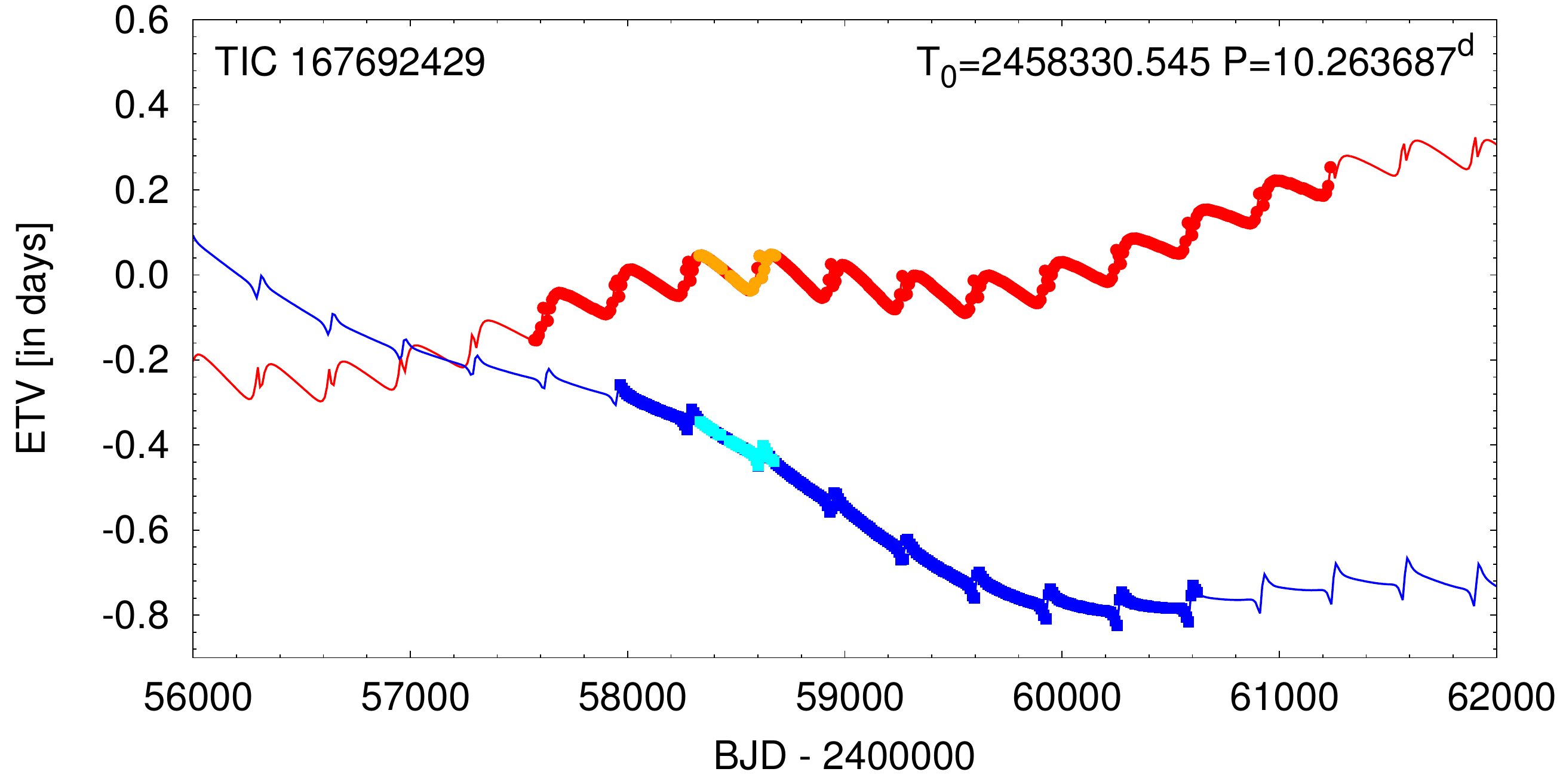}\includegraphics[width=0.49 \textwidth]{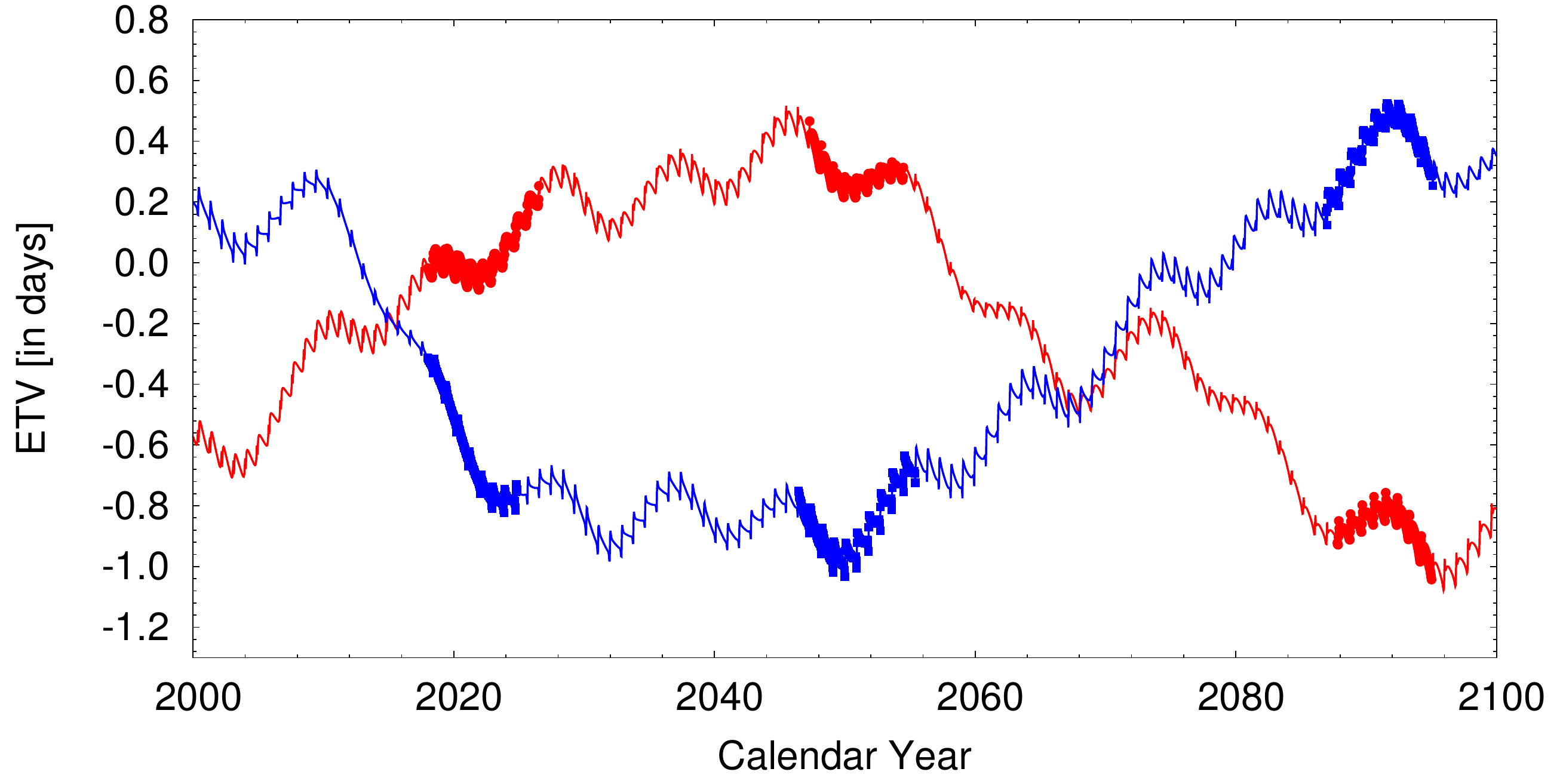}
\caption{Eclipse and conjunction timing variations of the inner orbit of TIC\,167692429 according to our photodynamical model on different time-scales during the $21^\mathrm{st}$ century. Conjunction times are for intervals without detectable eclipses; red and blue lines represent the inferior and superior conjunctions of the secondary component, respectively. During intervals of regular eclipsing, these lines are overplotted with heavier symbols corresponding to the primary (red) and secondary (blue) ETVs, also derived from the same model. Furthermore, in the left panel the observed ETVs calculated from the eclipse observations of the {\em TESS} spacecraft are also plotted with orange and cyan dots. Besides the dynamically forced apsidal motion, additional quasi-cyclic effect of other secular third-body perturbations are also distinicttly visible.}
\label{fig:T167692429ETVmidlong} 
\end{center}
\end{figure*}

\section{Summary and Conclusions}
\label{sec:disc}

In this work we have reported the discovery and complex analyses of the first two compact hierarchical triple star systems discovered with {\em TESS} in or near its southern continuous viewing zone during Year 1.  Both TICs\,167692429 and 220397947 were previously unknown EBs, and the presence of a third companion star was inferred from ETVs exhibiting signatures of strong 3rd-body perturbations and, in the first system, also from eclipse depth variations.  We carried out comprehensive analyses, including the simultaneous photodynamical modelling of {\em TESS} and archival ground-based WASP lightcurves, as well as ETV curves.  Also, for the first time, we included in the simultaneous fits multiple star SED data and theoretical PARSEC stellar isochrones, taking into account Gaia DR2 parallaxes and cataloged metallicities. 

TIC\,167692429 is found to be an eccentric triple star system consisting of two F-type twin stars forming the inner binary ($P_1=10\fd26$, $e_1=0.17$; $q_1=0.99$), while the third, less massive G-type star is on a moderately mutually inclined and eccentric orbit ($P_2=331\fd5$, $e_2=0.56$, $i_\mathrm{mut}=27\degr$; $q_2=0.34$). Given the mutually inclined configuration, the binary orbital plane precesses with a period of $P_\mathrm{apse}\approx70$\,yr, causing $\sim$10\,yr-long intervals where there are binary eclipses, interrupted by longer intervals with no binary eclipses, but during which irregular 3rd-body outer eclipses are predicted. We identify one likely outer eclipsing event near the end of WASP observations in 2012. 


In the absence of available radial velocity and quantitative spectroscopic observations, we used theoretical stellar isochrones and SED data for constraining effective temperatures, masses and other fundamental parameters of the stars being investigated. Our original idea was to obtain reliable stellar temperatures with a combination of (i) integrated SED information, (ii) photodynamically obtained mass ratios and relative stellar radii, and (iii) theoretical stellar isochrones. This process resulted in a multitude of stellar isochrones with different triplets of \{primary mass, age, metallicity\} which were found to be consistent with the SED and the photodynamical lightcurve solutions. Then, taking into account the accurate Gaia DR2 distances and auxiliary cataloged metallicities, we expected to find a proper narrow range of the appropriate isochrones which would be consistent with the astrometric distances to the systems.  In turn, this would lead to accurate stellar masses as well as stellar ages, metallicities, etc. Unfortunately, however, we found that either the Gaia distances or the metallicities, or both are strongly inconsistent with the solutions we obtained. 

For TIC\,167692429 the present Gaia DR2 distance would imply unphysically large stellar masses for the appropriate stellar radii, effective temperatures and metallicities. We interpret this inconsistency with the systematic effect of the almost 1-yr-period outer orbit on the trigonometric parallax measurements. As a consequence, in the present situation we are unable to obtain accurate stellar masses for this system and, furthermore, according to stellar isochrones TIC\,167692429 might be either a very young (pre-main sequence), or old (evolved, post-MS) system. If some future RV observations produce dynamical masses, one will be able to decide whether the post- or the pre-MS scenario is valid. Furthermore, sufficiently accurate dynamical masses could be used to determine the metallicity as well as a more accurate photometric distance. In addition, after obtaining very accurate Gaia DR3 distances and spectroscopically obtained dynamical masses, this triple would be appropriate for high-accuracy testing of stellar isochrones.

In the case of TIC\,220397947, this more compact coplanar triple has its binary formed by two F-type twins on an almost circular orbit ($P_1=3\fd55$, $e_1=0.001$; $q_1=0.95$), while the low-mass tertiary star has a rather short orbital period ($P_2=77\fd1$, $e_2=0.23$, $i_\mathrm{mut}=0\fdg6$; $q_2=0.25$). Archival WASP photometric observations reveal a discrepancy in the eclipse times which we interpret in terms of the presence of a fourth, low-mass star in the system with an orbital period of $P_3\approx2\,700$\,d. In the absence of radial velocity observations, we were unable to calculate accurate masses and ages for the two systems. According to stellar isochrones TIC\,167692429 might be either a very young (pre-main sequence), or old (evolved, post-MS) system. In the case of TIC\,220397947 our combined solution prefers a young, pre-MS scenario.

Both triples are currently scheduled to be observed during the {\em TESS} extended mission. Similar to the Year 1 measurements, TIC\,167692429 is likely  to be observed in all but one of the Year 3 sectors. Our photodynamical model predicts the deepest eclipses within the present 11\,yr-long cycle of regular eclipses during these observations.\footnote{Note, unfortunately, that the periastron passage of the outer orbit during that year and also the exact edge-on-view of the EB's orbital plane would occur right during Sector 35 measurements when the triple will not be visible to the {\em TESS} cameras.} This fact, combined with spectroscopic and RV measurements, not to mention the future orbital-motion-corrected Gaia DR3 results, should offer extraordinarily accurate fundamental stellar parameters and orbital elements for this system. Furthermore, regular monitoring of the normal binary eclipses over the next few years, up to the conclusion of the present cycle of regular eclipses, would also allow us to detect not only the dynamically forced apsidal motion, but also other kinds of secular three-body perturbations (see Fig.\ref{fig:T167692429ETVmidlong}). Note that the amplitude and (quasi-)period of these secular (or, in the present context one might say, `decadal') perturbations are very sensitive to the masses and the orbital configuration.  Therefore, an accurate detection of these features may also lead to extremely accurate masses and other dynamical parameters.

TIC\,220397947 is also expected to be re-observed in Year 3 sectors  29--33, 35, 36, 38, and 39. These observations, hopefully, will either verify or reject the quadruple system hypothesis. Independent of this, RV observations of this very tight, relatively bright, SB2 system would also offer the same advantages as in the case of our other system, making this triple or quadruple also a benchmark system for stellar evolutionary tracks and isochrones.

\section*{Acknowledgments}

T.\,B. acknowledges the financial support of the Hungarian National Research, Development and Innovation Office -- NKFIH Grant KH-130372.

This project has been supported by the Lend\"ulet Program of the Hungarian Academy of Sciences, project No. LP2018-7/2019 
and the MW-Gaia COST Action (CA 18104). 

This paper includes data collected by the {\em TESS} mission. Funding for the {\em TESS} mission is provided by the NASA Science Mission directorate. Some of the data presented in this paper were obtained from the Mikulski Archive for Space Telescopes (MAST). STScI is operated by the Association of Universities for Research in Astronomy, Inc., under NASA contract NAS5-26555. Support for MAST for non-HST data is provided by the NASA Office of Space Science via grant NNX09AF08G and by other grants and contracts.

This work has made use  of data  from the European  Space Agency (ESA)  mission {\it Gaia}\footnote{\url{https://www.cosmos.esa.int/gaia}},  processed  by  the {\it   Gaia}   Data   Processing   and  Analysis   Consortium   (DPAC, \url{https://www.cosmos.esa.int/web/gaia/dpac/consortium}).  Funding for the DPAC  has been provided  by national  institutions, in  particular the institutions participating in the {\it Gaia} Multilateral Agreement.

This publication makes use of data products from the Wide-field Infrared Survey Explorer, which is a joint project of the University of California, Los Angeles, and the Jet Propulsion Laboratory/California Institute of Technology, funded by the National Aeronautics and Space Administration. 

This publication makes use of data products from the Two Micron All Sky Survey, which is a joint project of the University of Massachusetts and the Infrared Processing and Analysis Center/California Institute of Technology, funded by the National Aeronautics and Space Administration and the National Science Foundation.

We  used the  Simbad  service  operated by  the  Centre des  Donn\'ees Stellaires (Strasbourg,  France) and the ESO  Science Archive Facility services (data  obtained under request number 396301).   





\end{document}